\documentclass{aa}
  \usepackage{graphicx}   

\begin{document}

  \title {Multicolor photometry of ten Seyfert 1 galaxies}

  \author{N. V. Boris\inst{1},  
	C. J. Donzelli\inst{1},
	M. G. Pastoriza\inst{2}\fnmsep\thanks{Visiting Astronomer, 
CTIO, NOAO, which is operated
  by AURA Inc., under contract to the NSF.},
	A. Rodriguez-Ardila$^{3,\star}$\fnmsep\thanks{On leave of absence 
        from Observatorio Astron\'omico Nacional, Universidad Nacional
        de Colombia, Bogot\'a},
\and
	D. L. Ferreiro\inst{1}}

	\offprints{C. J. Donzelli}

	\institute{IATE, Observatorio Astron\'omico, Universidad Nacional de
      C\'ordoba, Laprida 854, 5000, C\'ordoba, Argentina
	\and
	Departamento de Astronomia - UFRGS. Av. Bento Gon\c calves 9500,
  CEP 91501-970, Porto Alegre, RS, Brazil
	\and
	Instituto Astron\^omico e Geof\'{\i}sico - Universidade de S\~ao
  Paulo, Av. Miguel Stefano 4200, CEP 04301-904, S\~ao Paulo, SP, Brazil}

	\date{Received; accepted}

  \abstract{
  We present new valuable BVI photometry of ten Seyfert 1
  galaxies and narrow band H$\alpha$ images for six
  of these objects as well. The results indicate that the
  distribution of the luminosity of the sample has an amplitude
  of almost 4 magnitudes with an average of M$_B$=-20.7.
  The observed morphologies are confined to early type
  galaxies. A barred structure is found in only
  2 objects. Despite that early morphological types are
  dominant in this sample, integrated (B-V) colors are very blue.
  For instance, the SO galaxies show, on average, a
  (B-V)=0.78. This effect seems to be caused by the
  luminosity contribution of the active nucleus and/or
  the disk to the total luminosity of the galaxy. In
  the B band, the contribution of the active galactic
  nucleus to the total luminosity of the galaxy varies
  from 3\% to almost 60\% and the bulge to disk luminosity ratio
  (L$_{bulge}$/L$_{disk}$) ranges from 0.6 to 22.
  Signs of tidal interactions seems to be a common
  characteristic since they are observed in 6 of the
  objects and one of them seems to be located in a poor cluster
  not yet identified in the literature. 
  In contrast, H$\alpha$ extended emission is
  rare, with only 1 galaxy showing clear evidence of it.
  Luminosity profile decomposition shows that the model
  Gauss + bulge + disk properly reproduces the surface
  brightness of the galaxies. However,
  in order to account for the luminosity
  profile, most of the disk galaxies needs the inner 
  truncated exponential form with a central cutoff
  radius ranging from 3 to 10 kpc. This is interpreted in
  terms of reddened regions that are well identified in
  the B-V color maps. These regions present very similar
  colors among them, with (B-V)$\sim$1.2. This fact could be
  associated to the presence of dust confined in the
  inner regions of the galaxies.
	}

	\authorrunning{Boris et. al}
	\titlerunning{Photometry of Seyfert 1 galaxies}

	\maketitle

  \section{Introduction}
  The relationship between an active galactic nucleus (AGN)
  and its host galaxy is one of the key issues in the
  study of nuclear activity. One would expect that certain
  properties of active galaxies such as mass, luminosity,
  bulge to disk ratio and colors could influence nuclear
  activity or vice versa. In fact, the unified model for an
  AGN requires gas accretion, most probably from the host
  galaxy, onto a massive and compact object (Malkan 1983).
  Similarly, the nuclear burst scenario (Terlevich
  et al. 1992) requires gas fueling together with an
  efficient way to concentrate that gas into a small region
  of space, in a relatively short time-scale. Besides this,
  gas fueling seems to be connected to the presence of a bar,
  which is needed to get the non-axisymmetric potential that
  rises up from the theoretical works (Barnes \& Hernquist
  1991). However, recent observational studies based on
  optical data show that barred galaxies are not an specific
  signature for nuclear activity (Ho et al. 1997; Hunt et al.
  1999). In this sense, Regan \& Mulchaey (1999) based on the
  analysis of 12 Seyferts galaxies imaged with the HST
  propose central spiral dust lanes as an alternative
  mechanism to drive the gas to the central regions.

  Another important question to consider is the role played
  by the environment as a trigger of the nuclear activity.
  De Robertis, Hayhoe \& Yee (1998) have found that AGNs are
  not more likely to be associated to interactions than
  normal galaxies. However, Pastoriza, Donzelli \& Bonatto
  (1999), studying a sample of interacting galaxies, found
  that almost 40\% of the galaxies may host a low luminosity
  AGN.

  It is also worth noting that most of the studies on the
  topics exposed above are focused on the galaxy nuclei
  and some of them on the circumnuclear regions, but only a
  few works have paid attention to the properties of the
  stellar populations of the hosts galaxies. As an example,
  Gonz\'alez-Delgado et al. (1997)(hereafter GD97), presented H$\alpha$
  images of a sample of 55 active galaxies and S\'anchez-Portal
  et al. (2000), gave the results for broad band VRI and
  narrow band H$\alpha$ photometry for a sample of 24 nearby
  active galaxies. Also, Hunt et al. (1997) and M\'arquez et
  al. (2000) presented near infrared broad band images of a
  sample of 26 and 18 active galaxies respectively.\\ 
  Our goal in this paper is to present a new valuable set
  of photometric data for a sample of 10 Seyfert 1 galaxies
  and to describe the main properties of both the stellar
  component and the gas of the hosts galaxies. The paper is
  organized as follows: In Section 2 we describe the sample
  selection. In Section 3 we summarized the observations and
  data reduction. Section 4 discusses the photometric results
  and Section 5 describes the particular properties of each
  galaxy of the sample. Finally, in Section 6 are summarized
  our conclusions.

  \section{The sample}

  The galaxies chosen for this study were selected to span
  as broad a range of nuclear magnitudes and optical line
  characteristics as possible and still be classified as
  Seyfert 1 or Narrow Line Seyfert 1 as is the case for
  1H\,1934-063. For example, the Full-width half maximum of 
  the H$\alpha$ line derived from spectroscopy varies from 
  2000 km s$^{-1}$ to 7000 km s$^{-1}$ and the contribution
  of the stellar population to the observed continuum varies
  from negligible up to 80\% (Rodriguez-Ardila, 
  Pastoriza \& Donzelli 2000, hereafter RPD2000). These galaxies were selected
  from the compilation of many catalogs, mainly from the Calan Tololo Survey
  (Maza et al. 1989, 1992, 1994) and from the compilation of
  V\'eron-Cetty \& V\'eron (1996). However, due to limitations
  of the available observing time and the location of the
  observatory, the exact sample was a randomly selected
  subsample of our entire list of Seyfert 1. Those with little 
  or no photometric information available
  in the literature were chosen for this publication. 

  Figures 1a-j present our V images of the 
  sample galaxies. We have observed them to obtain photometric
  parameters such as isophotal shapes, colors and luminosity
  profiles that can provide information on their basic properties
  and structure. As a secondary objective, we want to contribute to
  the available multicolor photometric data for Seyfert 1 galaxies,
  that is still far to be completed.
  The continuum and line emission properties
  of six objects of our sample (CTS\,C16, CTS\,F10.01, CTS\,A08, 
  1H\,2107-097, CTS G03.04 and 1H\,1934-063) were already studied 
  by means of optical and near-IR spectroscopy (RPD2000).

   \begin{figure*}
   \centering
   \includegraphics[width=8.7cm,height=8.7cm]{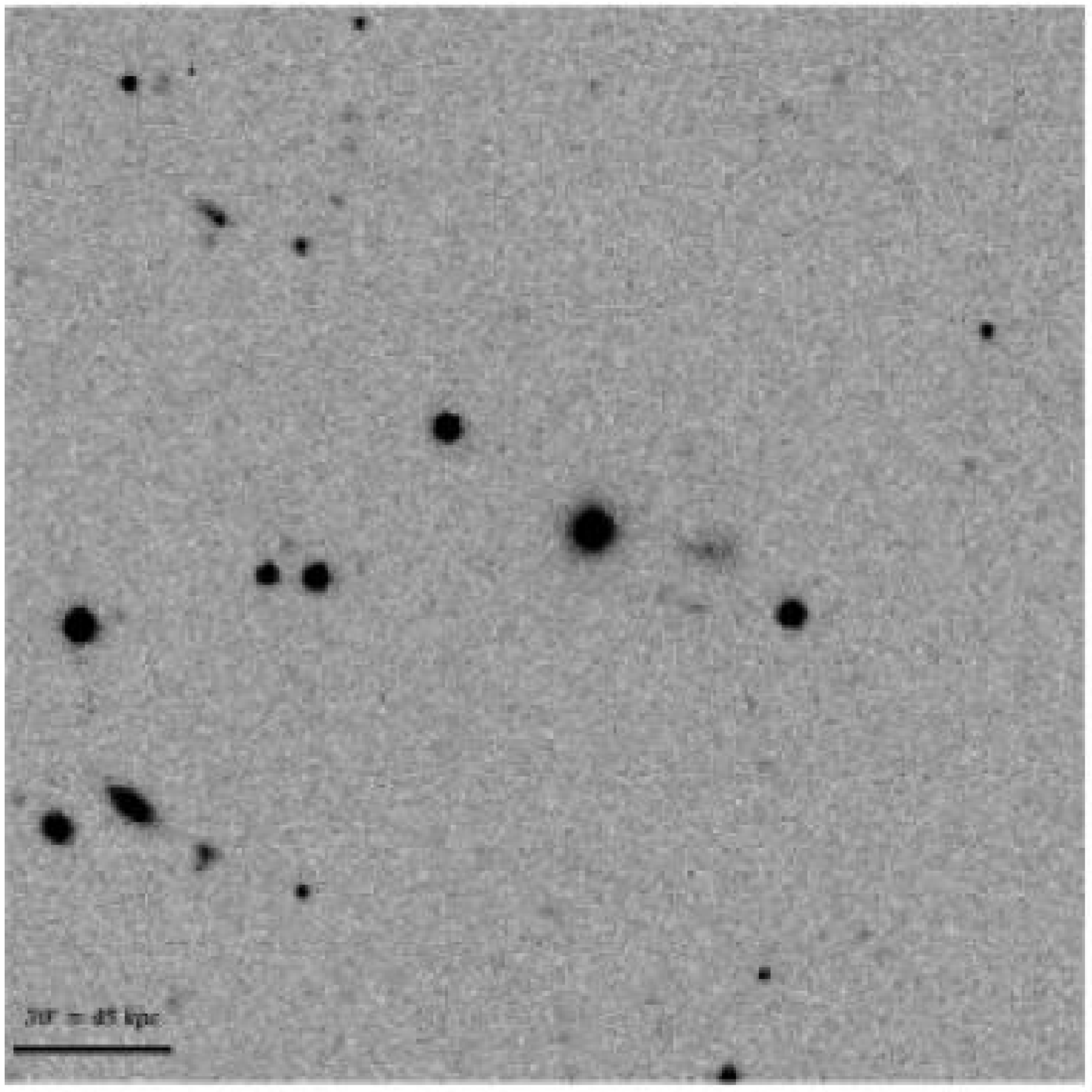}
   \includegraphics[width=8.7cm,height=8.7cm]{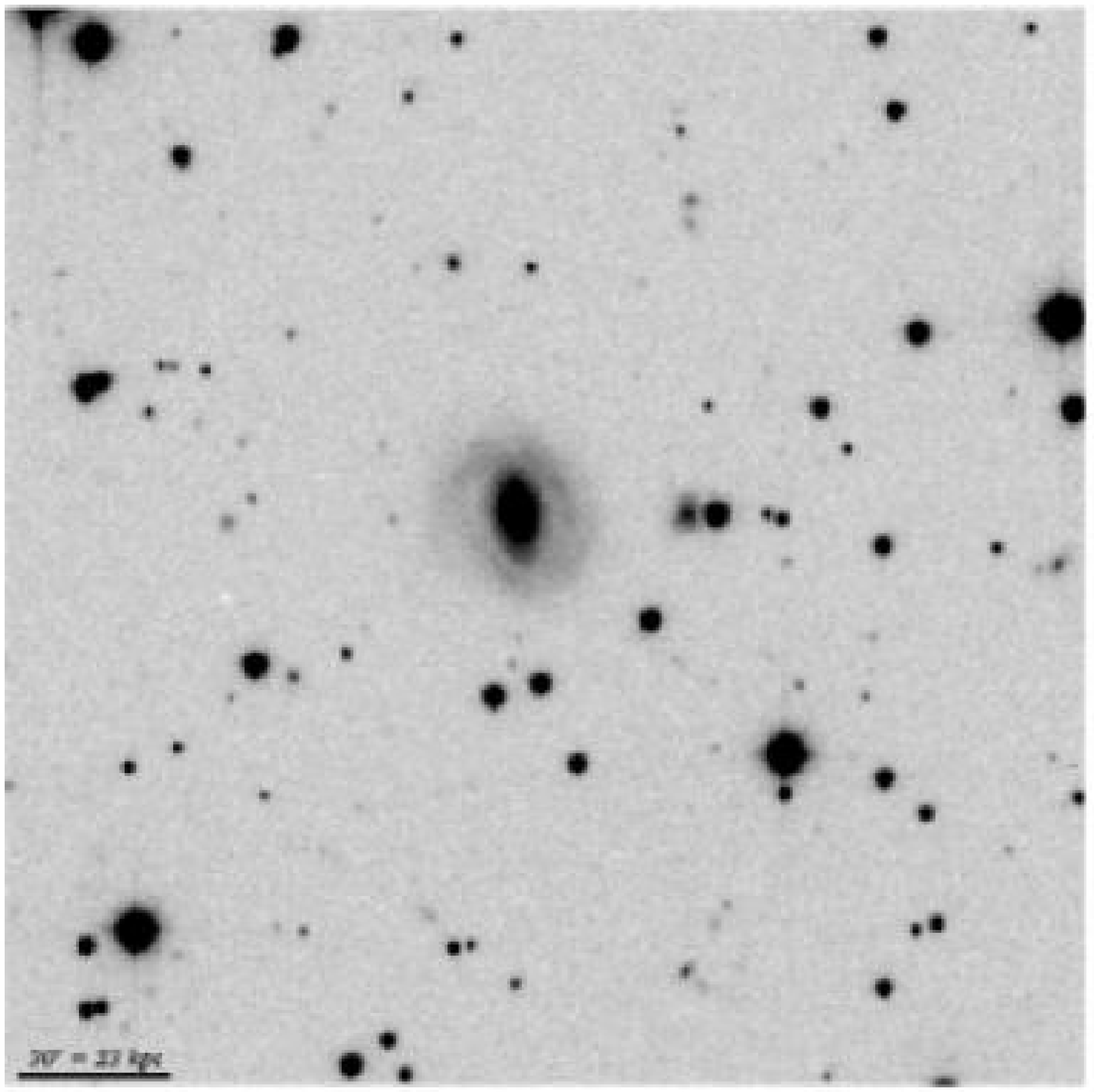}
   \includegraphics[width=8.7cm,height=8.7cm]{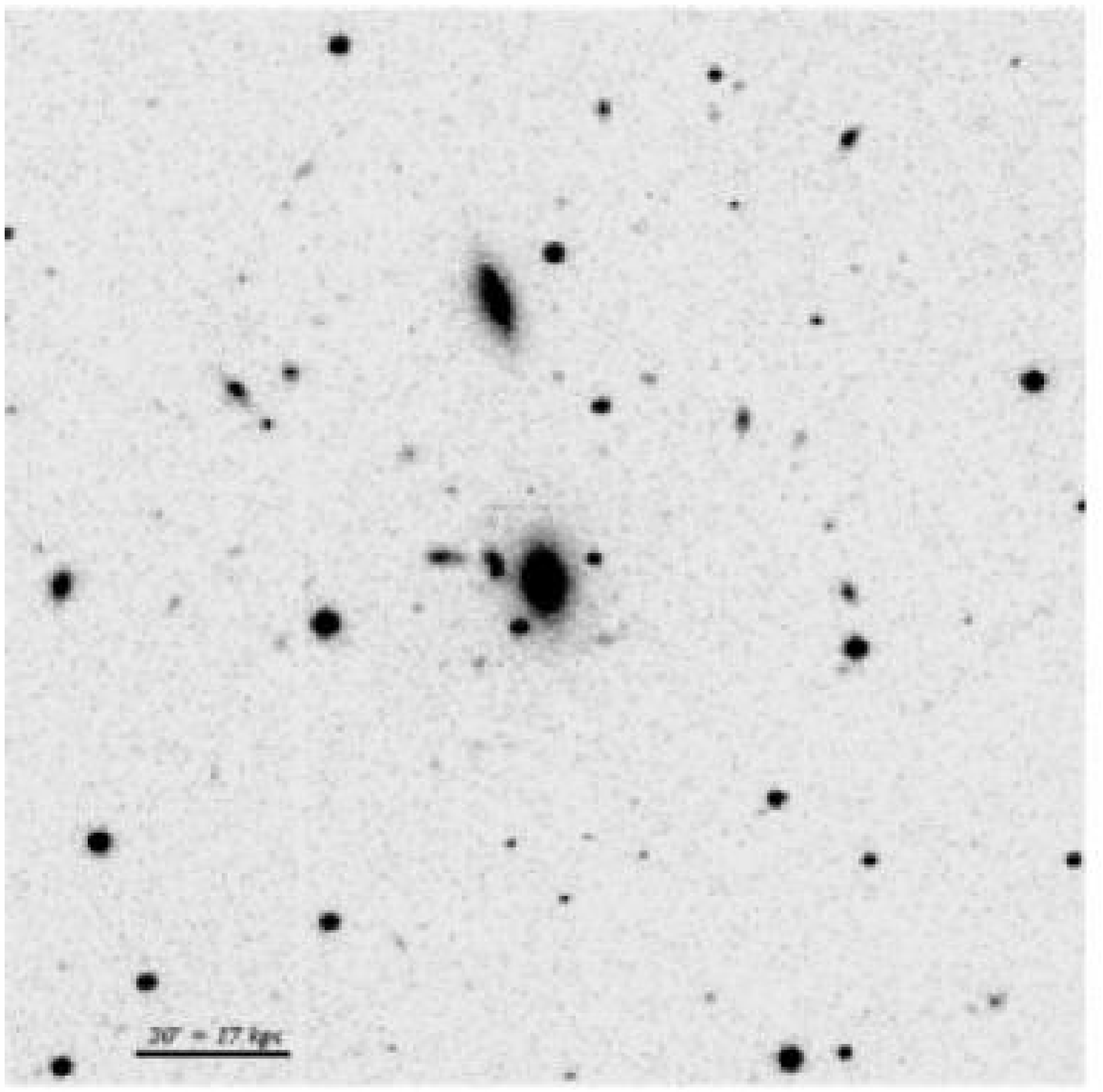}
   \includegraphics[width=8.7cm,height=8.7cm]{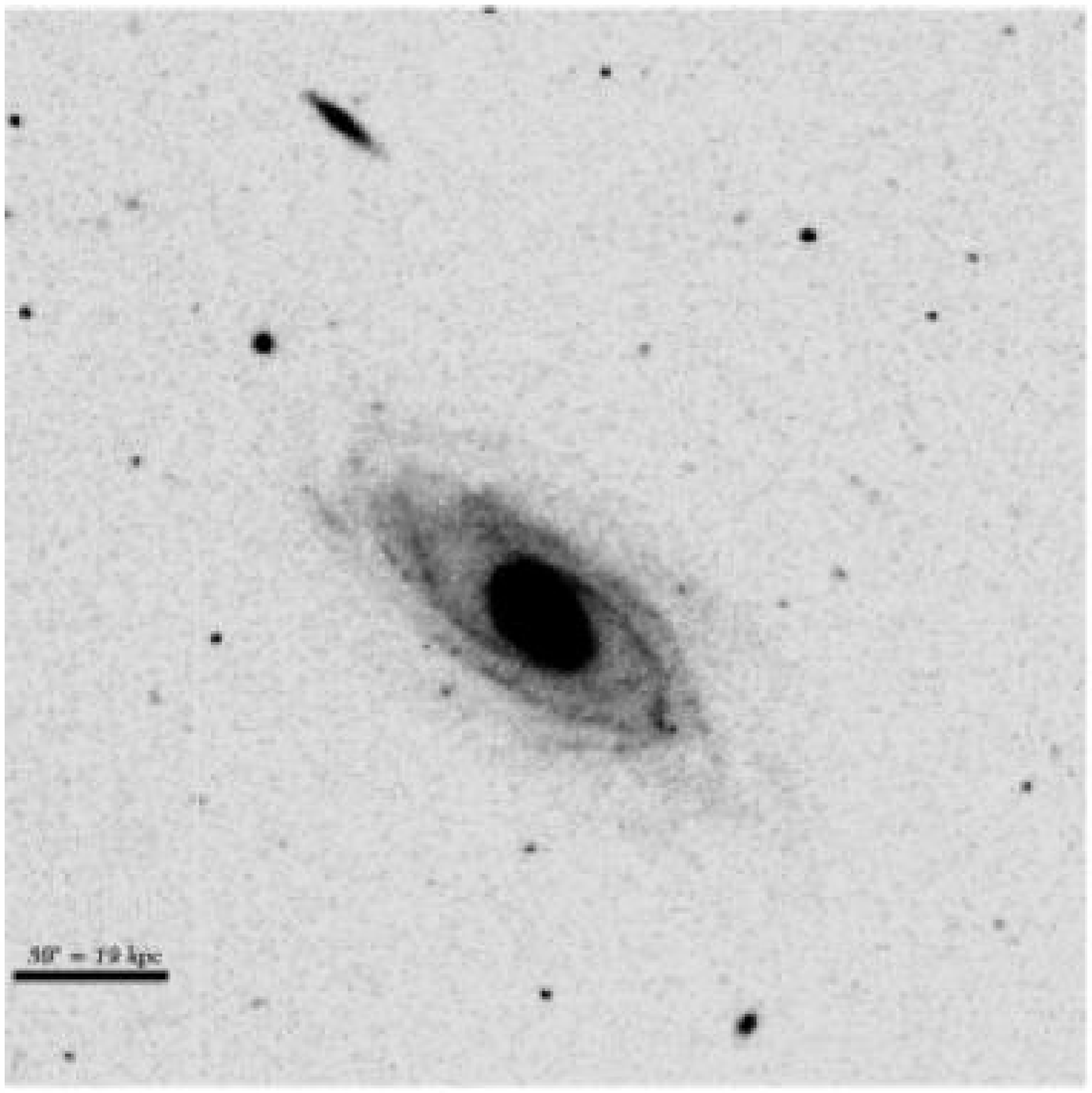}
   \includegraphics[width=8.7cm,height=8.7cm]{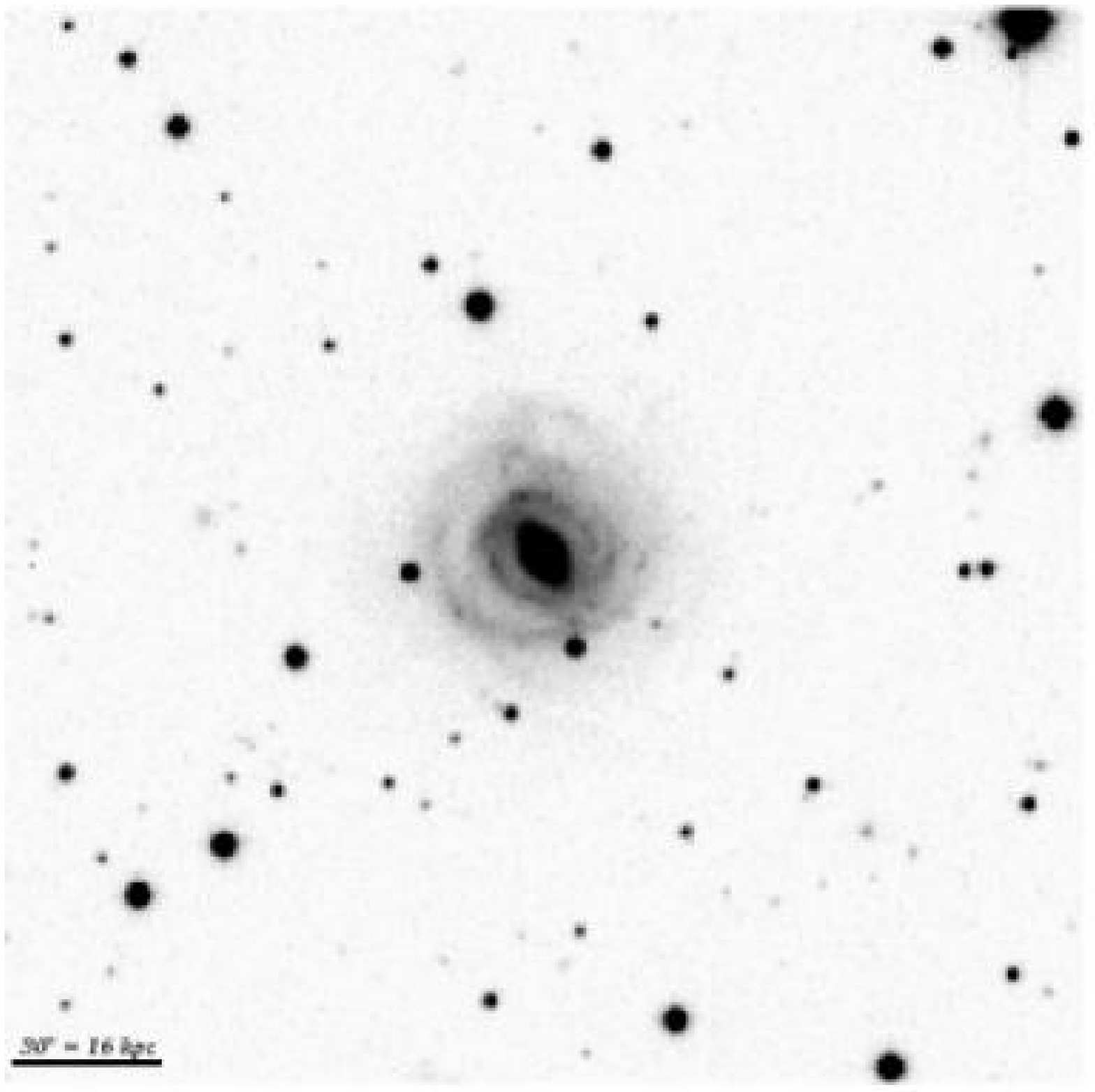}
   \includegraphics[width=8.7cm,height=8.7cm]{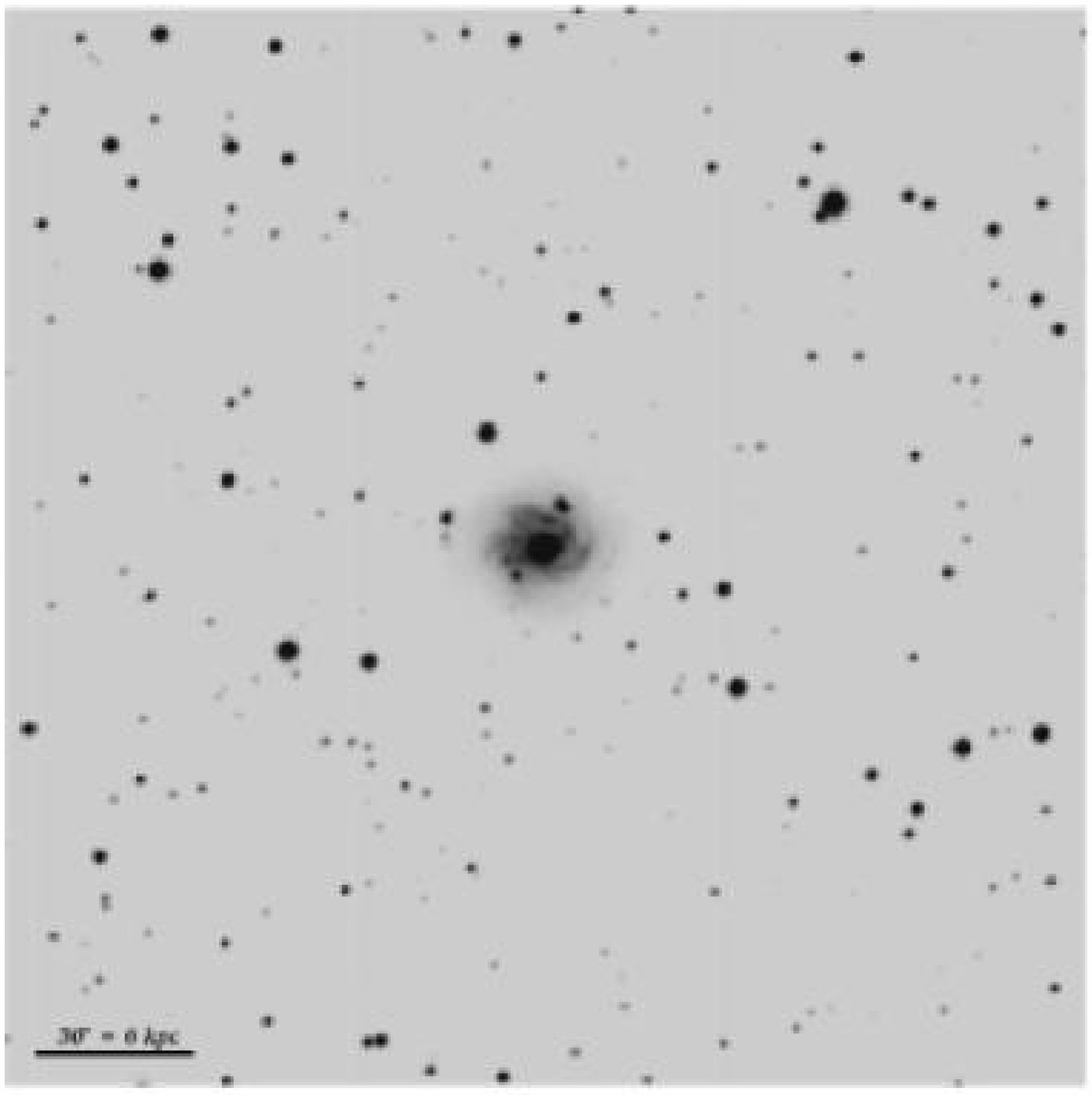}
	\end{figure*}

   \begin{figure*}
   \centering
   \includegraphics[width=8.7cm,height=8.7cm]{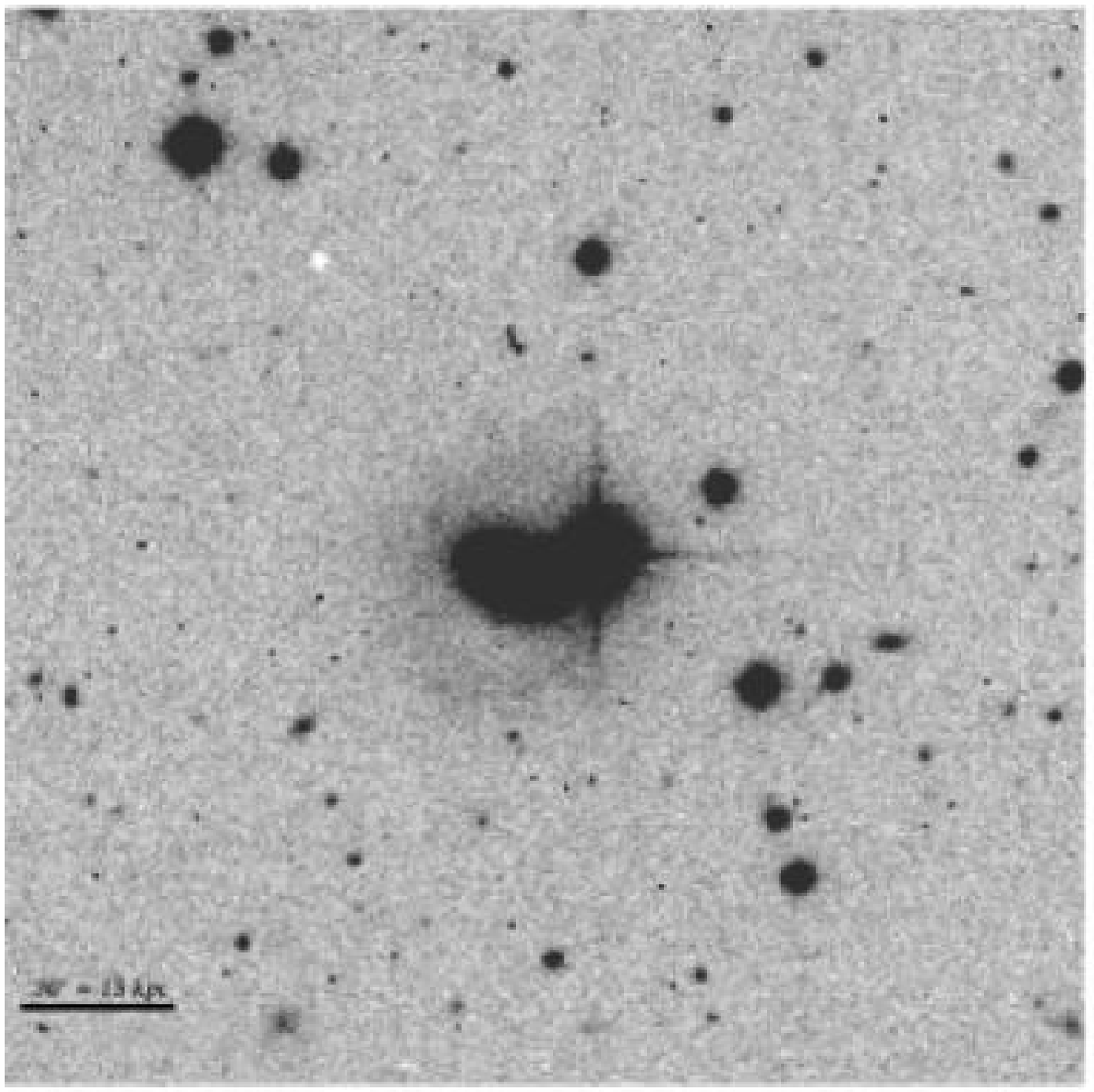}
   \includegraphics[width=8.7cm,height=8.7cm]{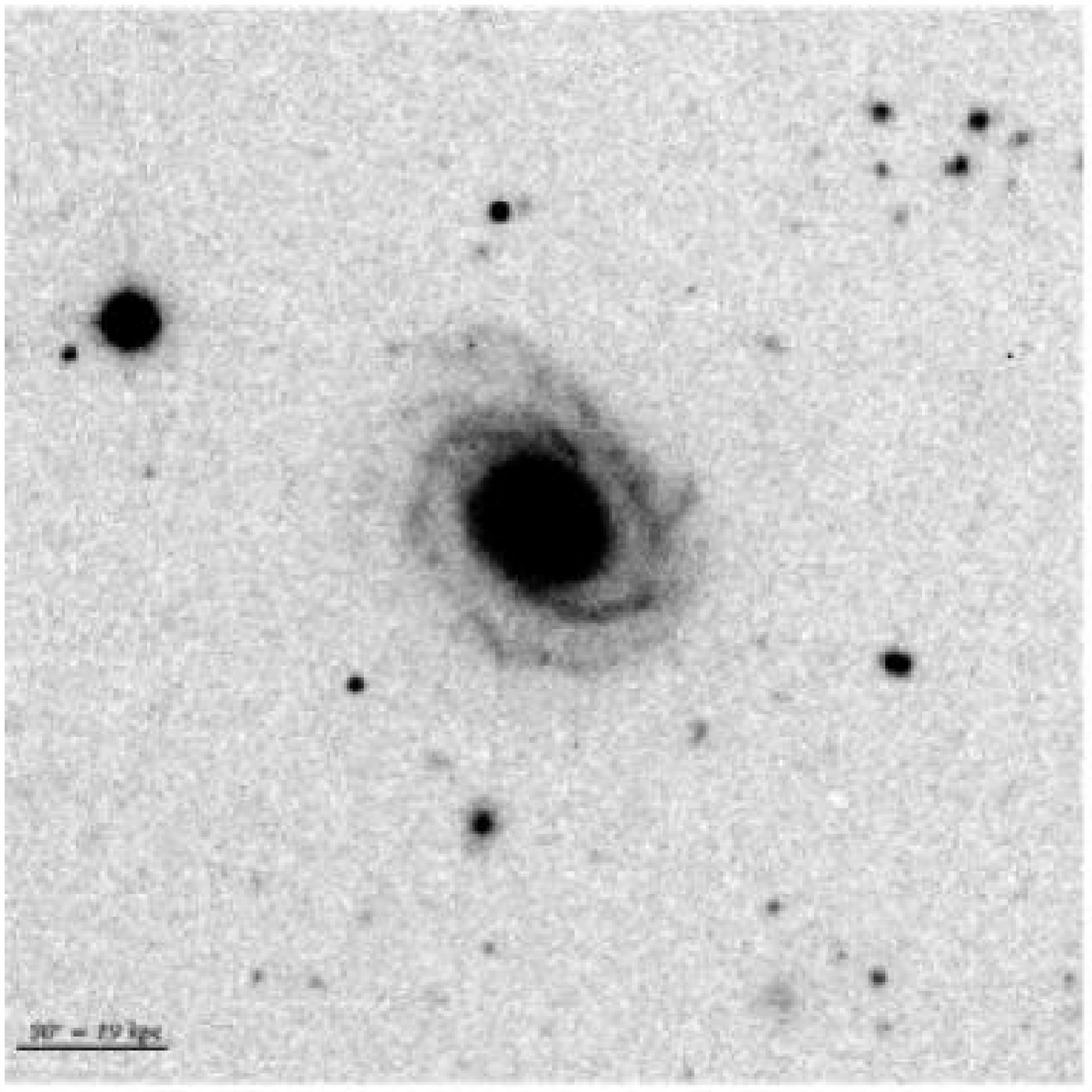}
   \includegraphics[width=8.7cm,height=8.7cm]{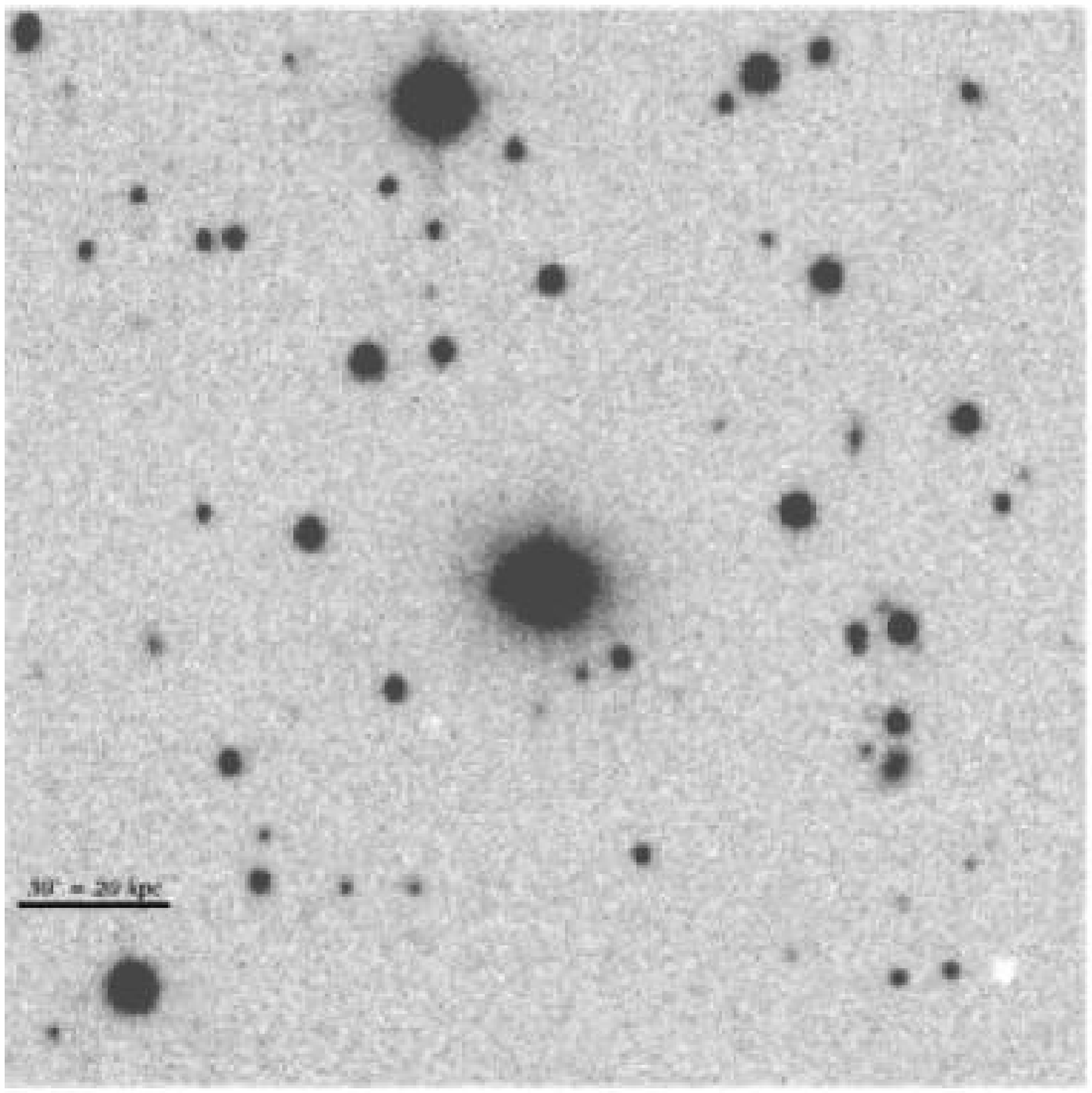}
   \includegraphics[width=8.7cm,height=8.7cm]{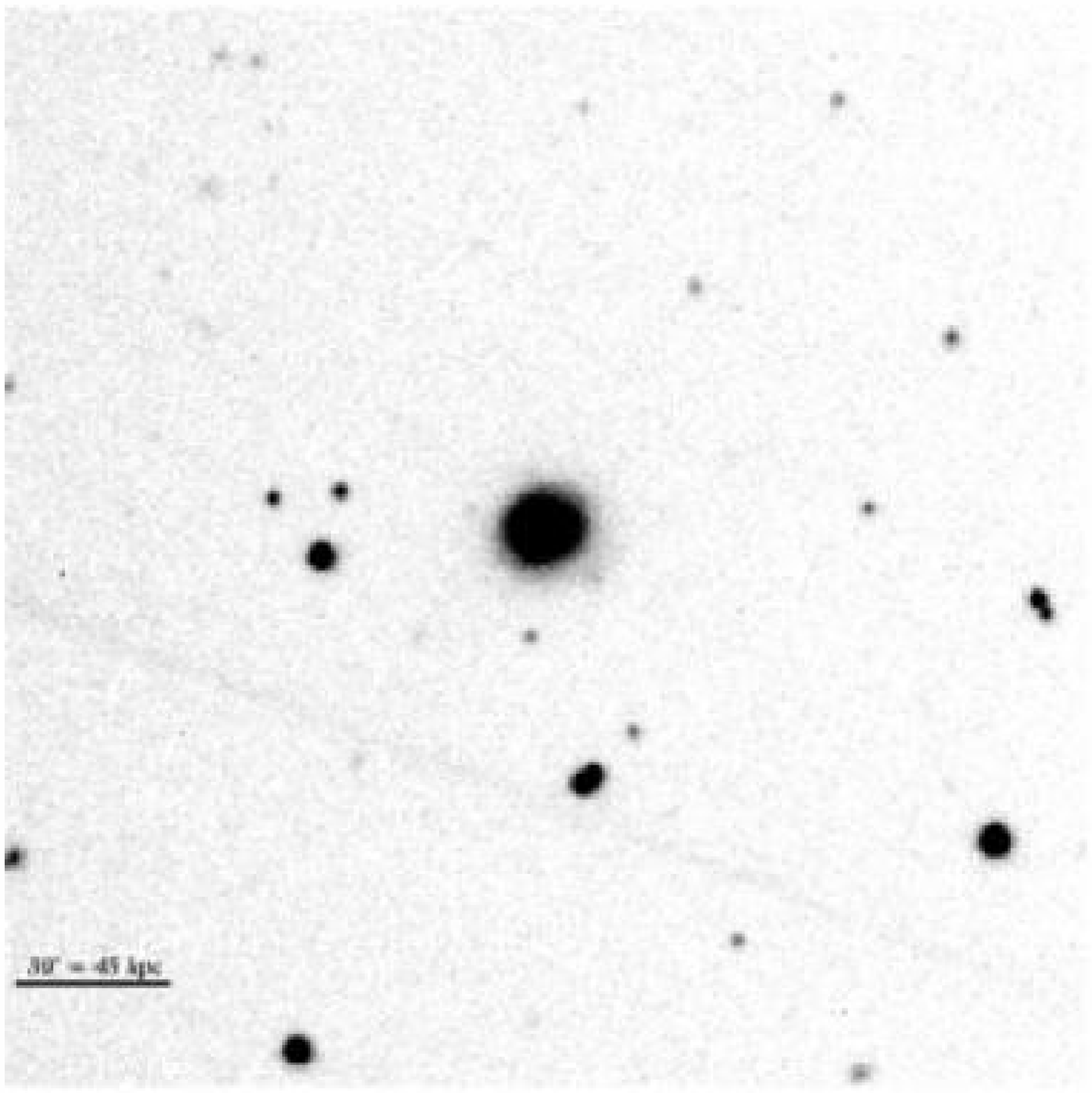}
   \caption{V images of the sample galaxies. North is on top and East to
  	the left. The lower left bar represents 30" and the corresponding projected scale in
	Kpc is also indicated.
  	a) CTC C16.16, b)CTS G03.04, c) CTS A08.12, d) ESO 602- G031,
    	e) ESO 025- G 002, f)1H 1934-063, g) 1H 2107-097, h) ESO 354- G 004, i) MRK
  	0509, j) CTS F10.01.}
    \end{figure*}

  The  galaxies in our sample are listed in Table~\ref{tab1} along with
  information about their morphological type, magnitudes, extinction 
  in blue band and radial velocities. These parameter were taken from 
  the NED\footnote{The NASA/IPAC Extragalactic Database (NED) is
  operated by the Jet Propulsion Laboratory, California Institute
  of Technology, under contract with the National Aeronautics and
  Space Administration.} database. Throughout this paper we have 
  adopted a Hubble constant $H_0$ = 75 km s$^{-1}$ Mpc$^{-1}$.
  
\begin{table*}
\caption[]{Basic information of the galaxy sample}
\label{tab1}
$$
\begin{array}{lcccccc}
\hline
\noalign{\smallskip}
 & R.A & \alpha &  & cz &  &  \\
\noalign{\smallskip}
Galaxy & (J2000) & (J2000) & m_{b} & (km s^{-1}) & Morphology & A_{b} \\
\noalign{\smallskip}
\hline
\noalign{\smallskip}
CTS~C16.16     & 00~ 00~ 53.5 & -44~ 39~ 34 & 18.0 & 23825 & ...          & 0.06 \\
CTS~G03.04     & 19~ 38~ 04.5 & -51~ 09~ 47 & 15.2 & 12082 & (R)SB0/a     & 0.27 \\
CTS~A08.12     & 21~ 32~ 02.2 & -33~ 42~ 54 & 15.7 &  8780 & ...          & 0.23 \\
ESO~602-G~031  & 22~ 36~ 55.9 & -22~ 13~ 15 & 14.9 & 10101 & (R')SAB(rs)b & 0.14 \\
ESO~025-G~002  & 18~ 54~ 39.8 & -78~ 53~ 51 & 14.6 &  8540 & (R:)SAB(r)b  & 0.69 \\
1H~1934-063    & 19~ 37~ 33.2 & -06~ 13~ 05 & 14.1 &  3174 & E            & 1.26 \\
1H~2107-097    & 21~ 09~ 09.8 & -09~ 40~ 16 & 14.3 &  8034 & ...          & 1.00 \\
ESO~354-G~004  & 01~ 51~ 41.7 & -36~ 11~ 16 & 15.1 & 10033 & (R':)SA(rs)b & 0.07 \\
MRK~509        & 20~ 44~ 09.7 & -10~ 43~ 25 & 13.0 & 10312 & compact      & 0.25 \\
CTS~F10.01     & 22~ 55~ 51.9 & -29~ 44~ 16 & 15.6 & 23500 & ...          & 0.09 \\
\noalign{\smallskip}
\hline
\end{array}
$$
\end{table*}

  \section{Observations and data reduction}

  \subsection{Observations}
  The sample was observed using the TEK 2048 CCD attached to the 0.9 m
  telescope of CTIO during the course of 3 nights in August 1998. 
  The pixel scale, projected to the sky plane, is 0.396$\arcsec$. 
  Broad B, V, I and four narrow filters 75 \AA\ wide centered at 
  $\lambda$ = 6477 \AA, 6649 \AA, 6781\AA\ and 7053\AA\ were employed.
  The first narrow filter was used to measure the continuum level
  adjacent to H$\alpha$ and the remaining three
  to isolate the H$\alpha$ + N[II] emission, depending on the 
  redshift of the object. Seeing conditions were very good
  (1.2$\arcsec$ to 1.5$\arcsec$) and all nights were photometric. 
  For every target object we obtained a set (usually 3 or 5) of short 
  (200 s - 400 s) exposures for each filter. A summary of the log
  of observations is shown in Table~\ref{logobs}.

\begin{table*}
\caption[]{Log. of Observations and exposure time in each
filter$^{\mathrm{a}}$} 
\label{logobs}
$$
\begin{array}{lcccccccc}
\hline
\noalign{\smallskip}
 &  &  &  &  & \multicolumn{4}{c}{H\alpha} \\
\cline{6-9}
\noalign{\smallskip}
Galaxy  & Date   & B  & V  & I & (\lambda 6477) & (\lambda 6649) & (\lambda 6781) & (\lambda 7053) \\
\noalign{\smallskip}
\hline
\noalign{\smallskip}
CTS~C16.16     & 15~ Aug~ 98 & 5x200 & 5x150 & 5x150 & ...   & ...   & ...   & ...\\
CTS~G03.04     & 17~ Aug~ 98 & 3x300 & 3x250 & 3x250 & ...   & ...   & ...   & ...\\
CTS~A08.12     & 15~ Aug~ 98 & 3x450 & 3x300 & 3x300 & 3x300 & ...   & 3x300 & ...\\
ESO~602-G~031  & 15~ Aug~ 98 & 3x450 & 3x350 & 3x300 & 3x300 & ...   & 3x300 & ...\\
ESO~025-G~002  & 15~ Aug~ 98 & 3x200 & 3x150 & 3x150 & 4x150 & ...   & 4x150 & ...\\
1H~1934-063    & 15~ Aug~ 98 & 3x400 & 3x250 & ...   & 3x250 & 3x250 & ...   & ...\\
1H~2107-097    & 17~ Aug~ 98 & 3x350 & 3x300 & 3x300 & ...   & ...   & ...   & ...\\ 
ESO~354-G~004  & 16~ Aug~ 98 & 4x400 & 3x300 & 3x300 & ...   & ...   & ...   & ...\\
MRK~509        & 16~ Aug~ 98 & 3x300 & 3x200 & 3x200 & 3x250 & ...   & 3x250 & ...\\
CTS~F10.01     & 16~ Aug~ 98 & 3x400 & 3x300 & 3x300 & 3x350 & ...   & ...   & 3x350\\
\noalign{\smallskip}
\hline
\end{array}
$$
\begin{list}{}{}
\item[$^{\mathrm{a}}$] For each filter we list the total number of frames 
multiplied by the exposure time of each frame, in seconds.
\end{list}
\end{table*}

  \subsection{Reduction of the BVI and H$\alpha$ images}
  The images were corrected for bias and flat-field  using 
  standard IRAF routines\footnote{IRAF is distributed by NOAO, which 
  is operated by AURA Inc., under contract to the NSF.}. Sky subtraction 
  was performed by averaging mean values of the sky, well beyond the 
  galaxy limits, in several boxes on every frame.
  All images for a particular galaxy were aligned and combined to
  obtain single B, V, I and H$\alpha$ images. The alignment was
  performed using the IMALIGN task of IRAF with at least 7 field
  stars. The typical accuracy was better than 0.05 pixel. 
  The same technique was used to generate the B-V and B-I color images as well as 
  to subtract the continuum emission from the H$\alpha$ images.
  When combining frames, if the individual images had different seeing, 
  the ones with better seeing were convolved with a Gaussian function in 
  order to match the image with the poorer seeing. 

  The photometric calibration to the Johnson-Cousins system
  was made using 4 or 5 standard stars per night selected from 
  Landolt (1992). The usual equations to transform the $b$, $v$ and $i$
  instrumental magnitudes into the V, (B-V) and (V-I) standard magnitudes were used.
  Estimates of accuracy in the calibrations are $\pm$ 0.04 mag in V, $\pm$ 0.06 mag 
  in (B-V) and $\pm$0.06 mag in (V-I).

  \section{Results}

  \subsection{Integrated photometric parameters}

The morphological classification, based on the galaxy colors, and 
the calculated magnitudes and colors for the objects in the sample
are listed in Table~\ref{morfo}. The magnitudes were derived by two 
independent methods. The first one is the integration of the intensity 
pixels in a series of diafragms with increasing radius until 
the sum converges. The second method consisted 
of the integration of the luminosity profile (see
\S~\ref{lumpro}). The results following the two
methods are in very good agreement. However, 
the comparison with the magnitudes listed in
The Third Reference Catalog of Bright Galaxies (de Vaucouleurs et al. 1991;
hereafter RC3), are in a reasonable agreement only for ESO\,602-G\, 031,
ESO\,025-G\,002, 1H\,1934-063, 1H\,2107-097, and  CTS\,F10.01. The remaining
objects show differences up to 0.5 mag. This is the case for MRK\,509 for 
which the RC3 lists m$_b$ = 13.0 while our value is m$_b$ = 13.50. 
Moreover, Kotilainen \& Ward (1994; hereafter KW94) 
obtained for the same galaxy m$_b$ = 14.06. We attribute the discrepancies 
to variability of the AGN.

\begin{table*}
\caption[]{Morphologies, magnitudes and colors derived for the galaxy sample} 
\label{morfo}
$$
\begin{array}{lcccccc}
\hline
\noalign{\smallskip}
Galaxy & Morphology^\mathrm{a} & D\times d^\mathrm{b} & M_{b} & B &
(B-V) & (V-I) \\
\noalign{\smallskip}
\hline
\noalign{\smallskip}
CTS~C16.16     & E1 & 5.5x5.5 & -20.36 & 17.32 & 0.91 & 0.79 \\
CTS~G03.04     & SO & 17.4x17.2 & -20.70 & 15.46 & 0.80 & 0.88 \\
CTS~A08.12     & SO & 9.5x9.2 & -19.24 & 16.23 & 1.01 & 0.94 \\
ESO~602-G~031  & SBa & 40.8x34.1 & -21.28 & 14.76 & 0.71 & 0.98 \\
ESO~025-G~002  & SBa & 26.1x26.1 & -20.85 & 14.67 & 0.85 & 1.36 \\
1H~1934-063    & Sb & 19.8x19.4 & -18.74 & 14.20 & 0.61 & ...   \\
1H~2107-097    & SO & 11.1x10.2 & -20.68 & 14.58 & 0.75 & 1.26 \\
ESO~354-G~004  & Sa & 34.4x34.1 & -21.22 & 14.65 & 0.88 & 1.13 \\
MRK~509        & SO? & 19.0x18.8 & -22.34 & 13.50 & 0.15 & 0.80 \\
CTS~F10.01     & SO & 12.3x12.2 & -21.99 & 15.60 & 0.67 & 0.86 \\
\noalign{\smallskip}
\hline
\end{array}
$$
\begin{list}{}{}
 \item[$^{\mathrm{a}}$] This paper.
 \item[$^{\mathrm{b}}$] Major and minor diameters in arcsec integrated up 
to m$_b$=25 mag arcsec$^{-2}$.
\end{list}
\end{table*}

The luminosity for the objects in the sample ranges from -18.7 to -22.3 
with an average M$_b$ = -20.7. This value is rather similar to that 
found by Yee (1983), M$_b$ = -20.4, for a sample of Seyfert galaxies, but 
somewhat higher when compared
to M$_b$ = -20.0 found by KW94 for a sample of Seyfert 1 galaxies. On the other hand, 
Christensen (1975) found that for a sample of normal spiral galaxies 
M$_B$ = -19.7. Note that M$_b$ values given by the other authors
have been recalculated using $H_0$ = 75 km s$^{-1}$ Mpc$^{-1}$.

  \subsection{Luminosity profiles} \label{lumpro}

  Since our sample is composed of both elliptical and spiral
  galaxies, we have used two different methods in order to obtain
  surface brightness profiles. For elliptical galaxies, profiles
  were obtained using the ELLIPSE routine within STSDAS
  (Jedrezejewski, 1987). Basically the task starts from a first
  guess elliptical isophote defined by approximate values for the
  center coordinates, ellipticity and position angle. With this
  initial values the image is sampled along an elliptical path
  producing an intensity distribution as a function of the
  position angle. Then the harmonic content of this
  distribution is analyzed by least-squares. The harmonic
  amplitudes together with the local image radial gradient are
  related to a specific ellipse geometric parameter and give
  information on how much the current parameter value deviates
  from the true one. The parameter is then modified by
  the calculated value and the process continues until
  convergence is reached.

  The approach was different for spiral galaxies because the
  ELLIPSE  algorithm does not converge due to the clumpy structure
  present in the spiral arms. In this case we have used the
  equivalent profile $m$ versus $r_{eq}$, where $r_{eq}$ = ($S/\pi$)$^{1/2}$
  being $S(m)$ proportional to the area projected on the image
  (in square arcsecs) subtended by all those points of the galaxy for
  which the intensity $I(m')$ = 10$^{-0.4m}$ satisfies the
  condition $I(m')> I(m)$  (S\'ersic, 1982). It is worth noting
  that this last profile and that used for an elliptical galaxy
  show exactly the same behavior for an E0 galaxy.

  The B, V and I profiles were then decomposed into 3 assumed
  components: Gaussian (due to the stellar-like profile of the 
  AGN), bulge and disk. In three cases we noted the presence of 
  other components such as bars, arms or lens. These structures 
  have not been taken into account in the fit because their 
  contribution to the total luminosity of the galaxy were not greater 
  than 5\%. The functional form adopted for each of the fitting 
  component is as follows:

  \begin{equation} \label{eq1}
  I(r)=I_0 exp(-2.71*(r/fw)^2)
  \end{equation}
  for the gaussian;
  \begin{equation} \label{eq2}
  I(r)=I_e exp(-7.688*((r/r_{eff})^{.25}-1))
  \end{equation} 
  for the bulge;
  \begin{equation} \label{eq3}
  I(r)=I_d exp(-r/d_l-(h_d/r)^3)
  \end{equation}
  for the disk component

  In the above expressions I$_0$ is the peak of the
  Gaussian profile at $r$=0 and $fw$ the full width at half maximum
  (FWHM). The quantity I$_e$ is the intensity
  at $r_{eff}$, the radius that encloses half of the total luminosity 
  of the bulge (also known as the effective radius). Finally, I$_d$ 
  is the central intensity, $d_l$ the length scale and $h_d$ the 
  radius of central cutoff of the disk component.

  In order to obtain the above parameters we followed the
  method described by Shombert \& Bothum (1987) using the NFIT
  routine implemented in STSDAS. This routine must be provided with
  appropriate initial parameters in order to begin the fit. Disk
  parameters can be guessed directly through the profile since
  the disk is not seriously contaminated by the bulge in the outermost
  region of the galaxy profile. However, it is necessary to have 
  photometric data at large radii in order to avoid contamination 
  from other more central components (Prieto et al. 1992). The main 
  difficulty was to perform the fit to the Gaussian and bulge 
  components since they are completely overlapped to each other. 
  This problem was solved by performing an initial fit considering only
  the innermost region data, generally the first 4$-$5$\arcsec$. During
  this process we fixed the disk guessed parameters and the $fw$ value 
  that was initially calculated using field stars. Task converges 
  rapidly at this step and calculated parameters do not significantly 
  depend on the initial adopted values. Finally, the calculated 
  parameters for the Gaussian and bulge components together with
  those of the disk were used as initial values to perform 
  the fit over the whole range of the luminosity profile, omitting 
  those points affected by additional structures. Uncertainties in
  the parameters were calculated by doing small variations to the 
  initial adopted values prior to the fitting.
  We found that the differences were never greater than 20\%. We have 
  also checked how seeing influences the calculated parameters by 
  deconvolving the images. The most seriously affected parameters are 
  that of the bulge ($I_0$ tends to be higher and $r_e$ tends to be 
  smaller) but the variations were never greater than 10-15\%. Moreover, 
  as deconvolution is a conservative process, the luminosity ratios 
  between the components remain unchanged between the errors.

  From the derived photometric parameters we were then able to
  calculate the total luminosity for each of the 3 components by integrating
  equations~\ref{eq1},~\ref{eq2} and~\ref{eq3} as follows:

  \begin{equation}
  L=\int_0^{\infty}I(r)2\pi r dr
  \end{equation}

  This integration leads to the following results:

  \begin{equation}
  L_{AGN}= \frac {\pi I_{g0} fw^2}{2.71} 
  \end{equation}
  for the gaussian
  \begin{equation}
  L_{bulge}=7.21\pi I_e r_{eff}^2 
  \end{equation}
  for the bulge
  \begin{equation}
  L_{disk}=2\pi I_0 d_l^2 
  \end{equation}
  for the disk component when h$_{d}$=0.

  For the case h$_{d} \neq$0, the above integral have no analytic 
  solution, so a numerical integration was adopted.
  Total magnitudes obtained using the above procedure are, on average, 
  0.10 mag brighter than those calculated through the integration 
  of the image intensity pixels, as was explained in Section 4.1. This
  systematic difference is not surprising since now
  the luminosity profile integration is made up to r=$\infty$.
 
\begin{table*}
\caption{Photometric parameters derived for the galaxy sample}
\label{photpar}
$$
\begin{array}{lccccccccccccccccccccccccccc}
\hline
\noalign{\smallskip}
& \multicolumn{3}{c}{g_{\rm o}^\mathrm{a}} & & \multicolumn{3}{c}{f_{\rm w}^\mathrm{b}} & &
\multicolumn{3}{c}{m_{\rm e}^\mathrm{a}} & & \multicolumn{3}{c}{r_{\rm e}^\mathrm{c}} & &
\multicolumn{3}{c}{b_{\rm o}^\mathrm{a}} & & \multicolumn{3}{c}{d_{\rm l}^\mathrm{c}} & &
\multicolumn{3}{c}{h_{\rm d}^\mathrm{c}} \\
\cline{2-4}  \cline{6-8}  \cline{10-12}  \cline{14-16}  \cline{18-20}
\cline{22-24} \cline{26-28}
\noalign{\smallskip}
Galaxy & 
B & V & I & & B & V & I & & B & V & I & & B & V & I & &
B & V & I & & B & V & I & & B & V & I  \\
\noalign{\smallskip}
\hline
\noalign{\smallskip}
CTS~C16.16     & 20.1 & 19.8 & 19.2 & & 2.0  & 2.2  & 1.8  & & 21.3 & 20.3 &
19.5 & & 2.0  & 3.1  & 2.2  & & ...  & ...  & ...  & & ...  & ...  & ... 
&  & ...  & ...  & ... \\
CTS~G03.04     & 18.3 & 17.6 & 16.7 & & 1.5  & 1.5  & 1.3  & & 20.9 & 20.4 &
21.0 & & 1.2  & 1.6  & 3.7  & & 22.8 & 21.7 & 21.0 & & 6.8  & 5.6  & 5.2 
&  & ...  & ...  & ... \\
CTS~A08.12     & 18.2 & 17.5 & 17.7 & & 1.6  & 1.4  & 1.3  & & 20.8 & 19.7 &
18.4 & & 0.7  & 0.8  & 0.6  & & 24.5 & 23.9 & 22.4 & & 5.5  & 7.3  & 7.7  
&  & 3.4  & 2.4  & 1.4 \\
ESO~602-G~031 & 19.2 & 18.7 & 18.8 & & 1.6  & 1.8  & 1.7  & & 24.3 & 23.3 &
22.1 & & 11.5 & 9.8  & 8.7  & & 22.0 & 21.7 & 20.0 & & 6.2  & 6.2  & 6.2  
&  & 10.2 & 10.5 & 12.0 \\
ESO~025-G~002 & 20.1 & 19.7 & 18.6 & & 2.1  & 2.1  & 2.0  & & 23.0 & 22.2 &
20.7 & & 3.7  & 4.4  & 4.2  & & 21.9 & 20.8 & 19.8 & & 6.1  & 5.0  & 5.2
&  & 3.9  & 4.6  & 4.7 \\
1H~1934-063    & 17.8 & 18.1 & ...  & & 2.0  & 2.0  & ...  & & 24.3 & 20.7 &
...  & & 2.3  & 0.5  & ...  & & 20.4 & 20.1 & ...  & & 1.0  & 1.2  & ...  
&  & ...  & ...  & ... \\   
1H~2107-097    & 17.2 & 17.2 & 16.3 & & 1.9  & 1.8  & 1.7  & & 18.7 & 17.9 &
19.0 & & 0.6  & 0.6  & 1.6  & & 20.3 & 20.0 & 19.3 & & 1.8  & 2.3  & 3.3  & &
3.4  & 2.8  & 2.9 \\
ESO~354-G~004 & 19.8 & 18.8 & 17.6 & & 2.3  & 2.0  & 2.8  & & 21.7 & 20.6 & 
20.4 & & 3.3  & 3.2  & 4.9  & & 22.1 & 21.9 & 20.1 & & 6.6  & 6.6  & 4.4  & & 10.3 &
10.0 & 9.7 \\ 
MRK~509        & 16.5 & 16.7 & 16.6 & & 2.9  & 2.8  & 2.8  & & 18.3 & 18.2 &
17.1 & & 0.9  & 1.1  & 1.0  & & 21.7 & 20.8 & 19.6 & & 3.5  & 3.5  & 3.5  & &
8.7  & 8.7  & 8.7 \\
CTS~F10.01     & 19.1 & 18.4 & 17.9 & & 2.4  & 2.1  & 2.1  & & 19.7 & 19.3 &
18.2 & & 1.2  & 1.4  & 0.9  & & 20.7 & 20.1 & 19.0 & & 4.6  & 4.9  & 5.0  & & 
...  & ...  & ... \\
\noalign{\smallskip}
\hline
\end{array}
$$
\begin{list}{}{}
\item[$^{\mathrm{a}}$] In units of mag arcsec$^{-2}$
\item[$^{\mathrm{b}}$] In units of arcseconds ($\arcsec$)
\item[$^{\mathrm{c}}$] In units of kpc
\end{list}
\end{table*}

\begin{table*}
\caption{Luminosity ratios between components}
\label{fotdec}
$$
\begin{array}{lccccccccccc}
\hline
\noalign{\smallskip}
 & \multicolumn{3}{c}{L_{bulge}/L_{disk}} & &
\multicolumn{3}{c}{L_{bulge}/L_{gauss}} & &
\multicolumn{3}{c}{L_{gauss}/L_{total}}\\
\cline{2-4}  \cline{6-8}  \cline{10-12}
\noalign{\smallskip}
Galaxy &
B & V & I & & B & V & I & & B & V & I \\
\noalign{\smallskip}
\hline
\noalign{\smallskip}
CTS~C16.16     & ... & ... & ... & &  2.7 & 4.3 & 8.9 & &  0.27& 0.19& 0.10 \\
CTS~G03.04     & 0.6 & 1.0 & 1.8 & &  2.0 & 2.9 & 5.6 & &  0.16& 0.15& 0.10 \\
CTS~A08.12     & 2.0 & 2.0 & 1.0 & &  1.2 & 2.5 & 7.6 & &  0.36& 0.21& 0.06 \\
ESO~602-G~031  & 3.9 & 6.1 & 3.3 & &  43.0& 20.7& 56.0& &  0.02& 0.04& 0.01 \\
ESO~025-G~002  & 0.6 & 1.1 & 1.5 & &  13.5& 27.2& 41.5& &  0.03& 0.02& 0.01 \\
1H~1934-063    & 0.6 & 0.4 & ... & &  1.6 & 3.2 & ... & &  0.18& 0.09& ...  \\
1H~2107-097    & 4.2 & 3.2 & 1.7 & &  1.7 & 4.4 & 5.5 & &  0.33& 0.15& 0.10 \\
ESO~354-G~004  & 3.0 & 6.2 & 11.9& &  16.8& 22.5& 10.9& &  0.04& 0.04& 0.08 \\
MRK~509        & 22.1& 16.1& 7.1 & &  0.8 & 1.6 & 3.5 & &  0.56& 0.37& 0.20 \\
CTS~F10.01     & 0.6 & 0.6 & 0.3 & &  1.2 & 1.6 & 1.2 & &  0.24& 0.19& 0.14 \\
\noalign{\smallskip}
\hline
\end{array}
$$
\end{table*}

  Figures 2a-j present the observed luminosity profiles and the
  fit applied to each galaxy. Color profiles have been obtained
  directly from the fitted profiles. Decomposition in the three 
  components is only shown for the B data. Table~\ref{photpar} lists the 
  photometric parameters calculated from the decomposition technique 
  and Table~\ref{fotdec} lists the bulge to disk, 
  bulge to Gaussian and Gaussian 
  to total luminosity ratios in the B, V and I filters. 
  Note that in Table~\ref{photpar}, 
  g$_0$=-2.5 log(I$_0$), m$_e$=-2.5 log(I$_e$) 
  and b$_0$=-2.5 log(I$_d$). These quantities are expressed in 
  mag arcsec$^{-2}$.

A very interesting result was obtained from the decomposition profiles:
In six out of eight disk galaxies it was necesary to truncate
the exponential profile in order to get an adecuate fit. In fact, the 
radius of the central cutoff $h_d$ ranged from 3 kpc for CTS A08.12 and
1H 2107-097, up to 10 kpc as is the case of ESO 602-G031. Moreover, 
for each galaxy the cutoff radius corresponds to a reddened region that is well 
identified in the B-V color map. We have obtained very similar colors
for these regions, (B-V) $\sim$ 1.2. Color maps for three galaxies
of the sample are presented in \S~ 5. 

  \begin{figure*}
   \centering
   \includegraphics[width=8.7cm,height=8.7cm]{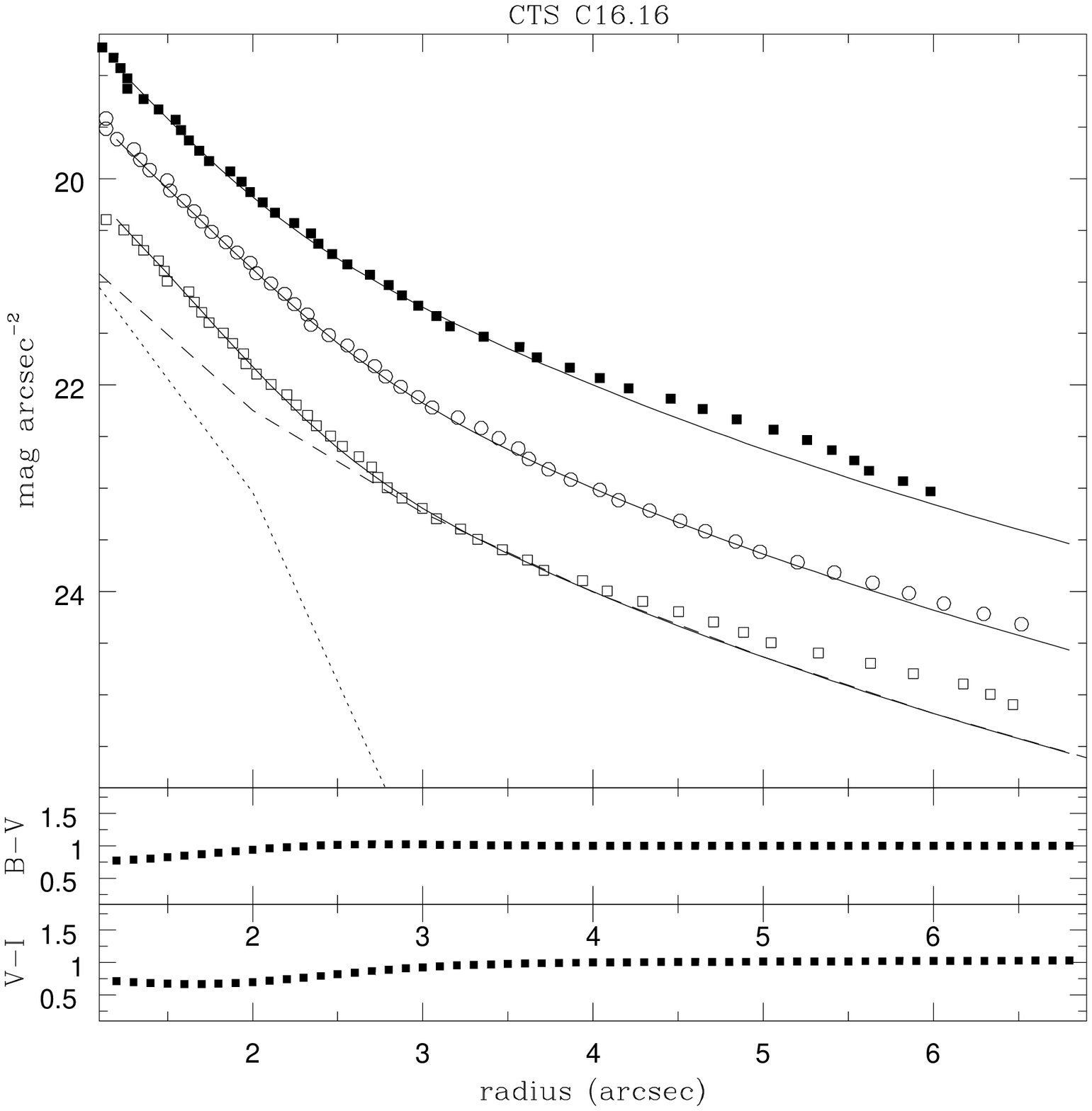}
   \includegraphics[width=8.7cm,height=8.7cm]{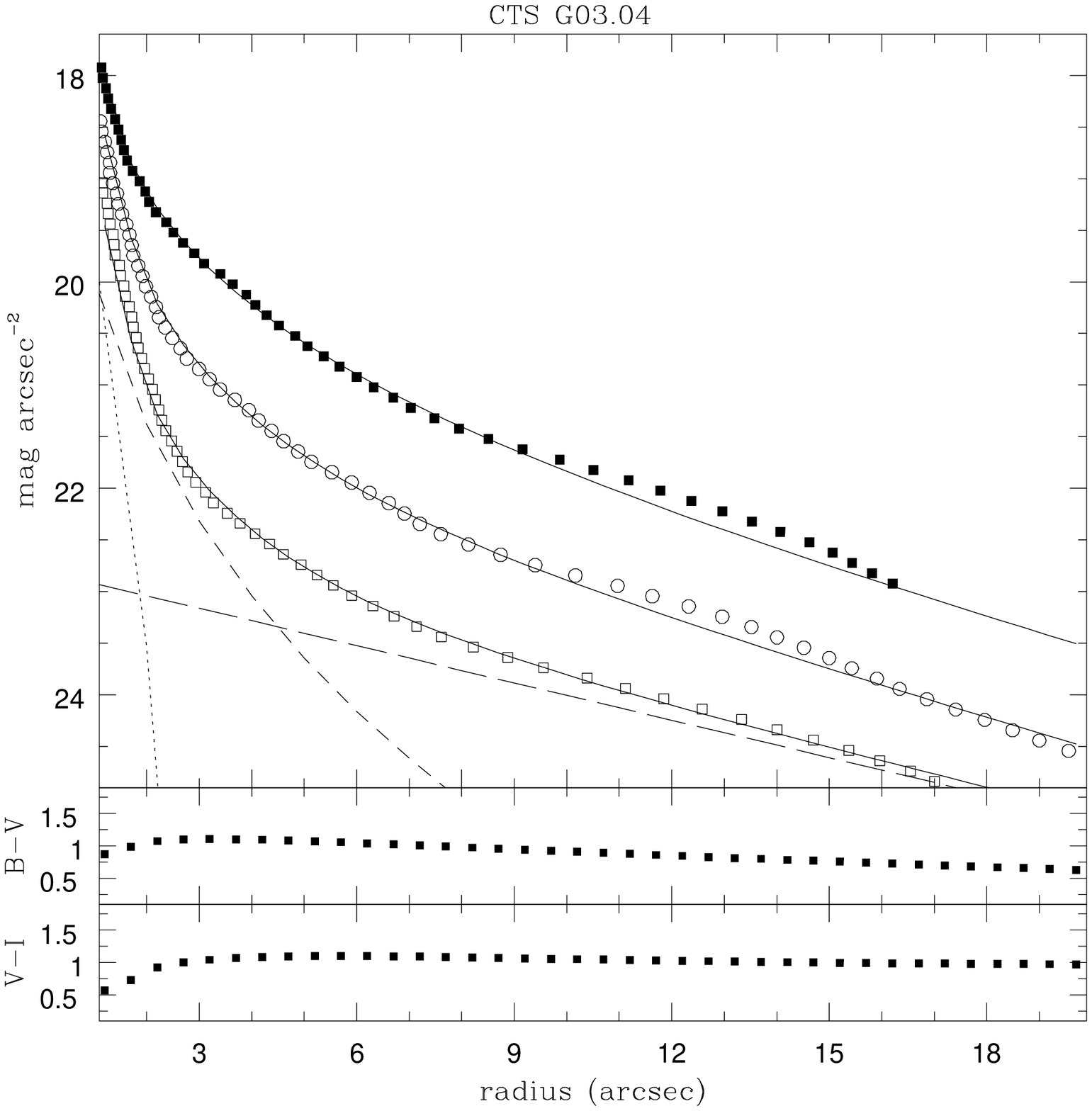}
   \includegraphics[width=8.7cm,height=8.7cm]{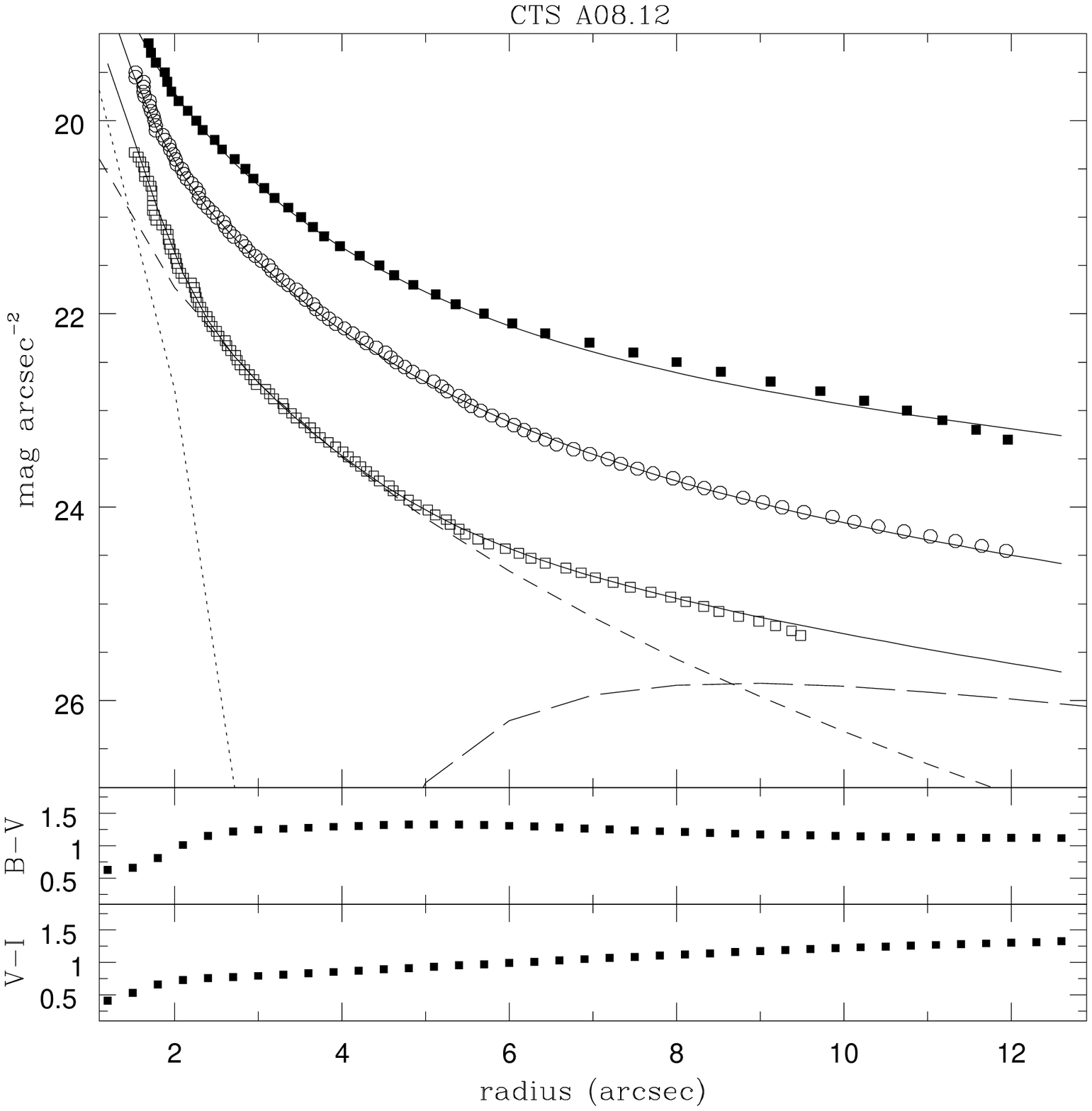}
  \includegraphics[width=8.7cm,height=8.7cm]{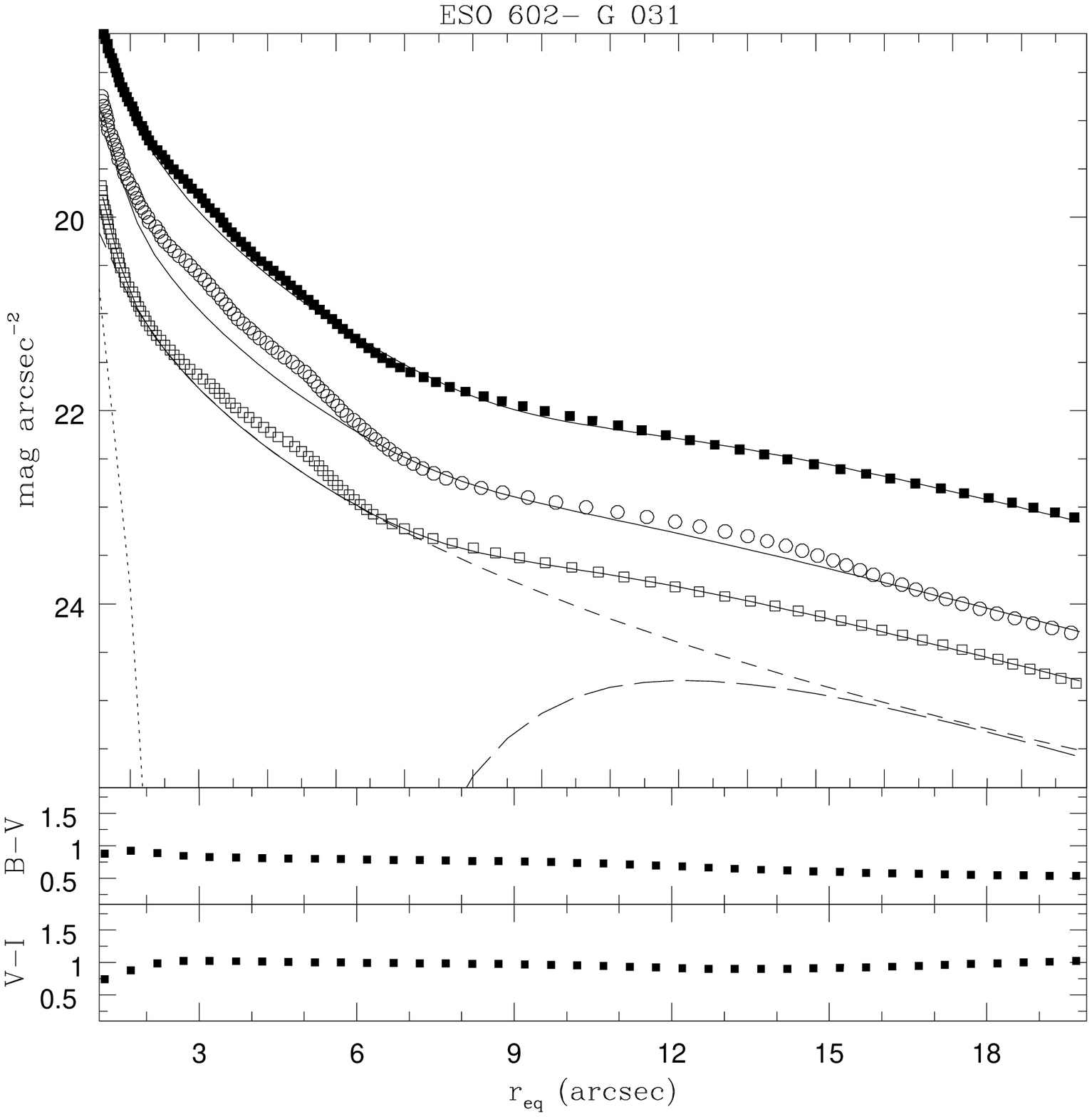}
   \includegraphics[width=8.7cm,height=8.7cm]{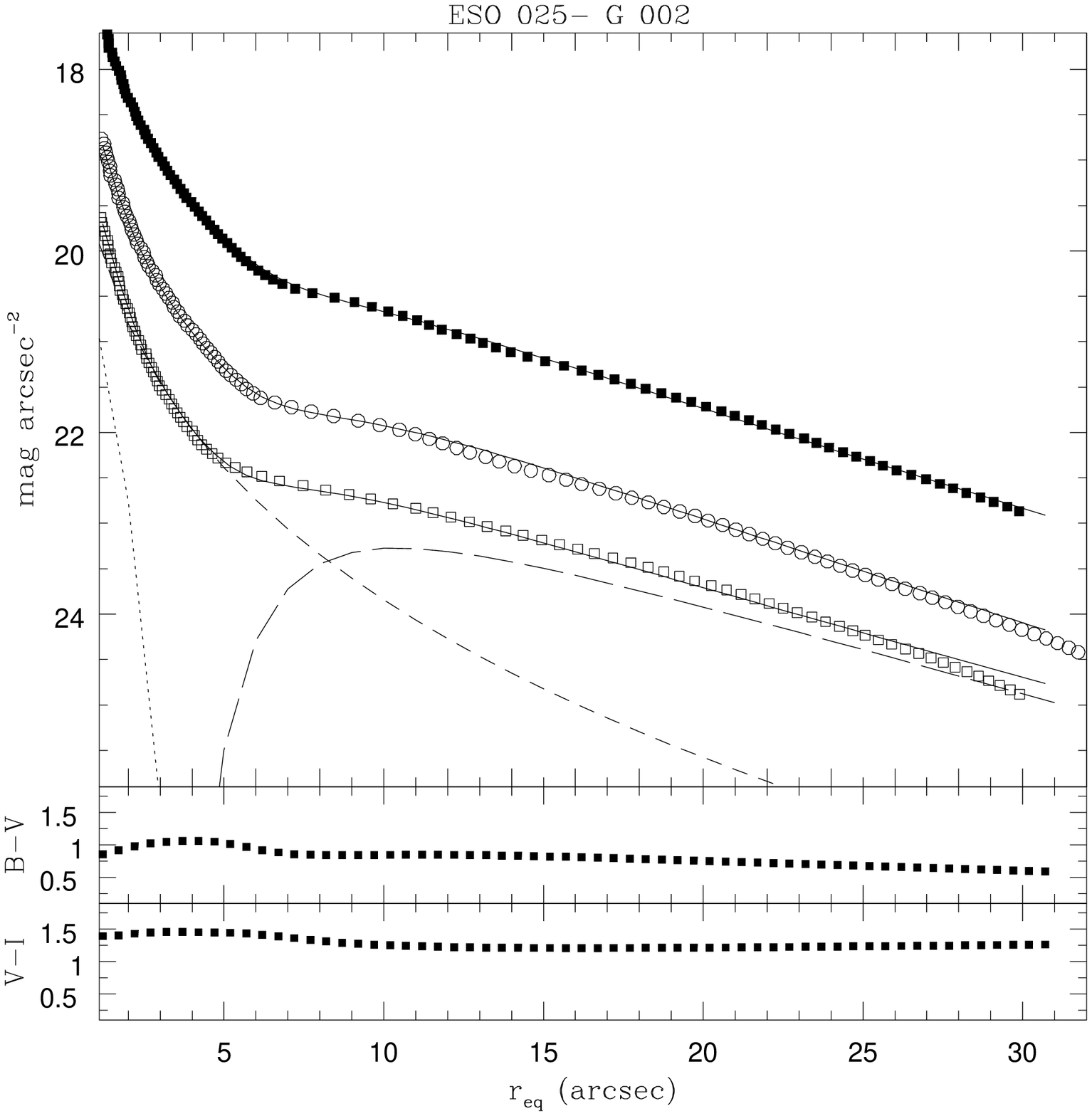}
   \includegraphics[width=8.7cm,height=8.7cm]{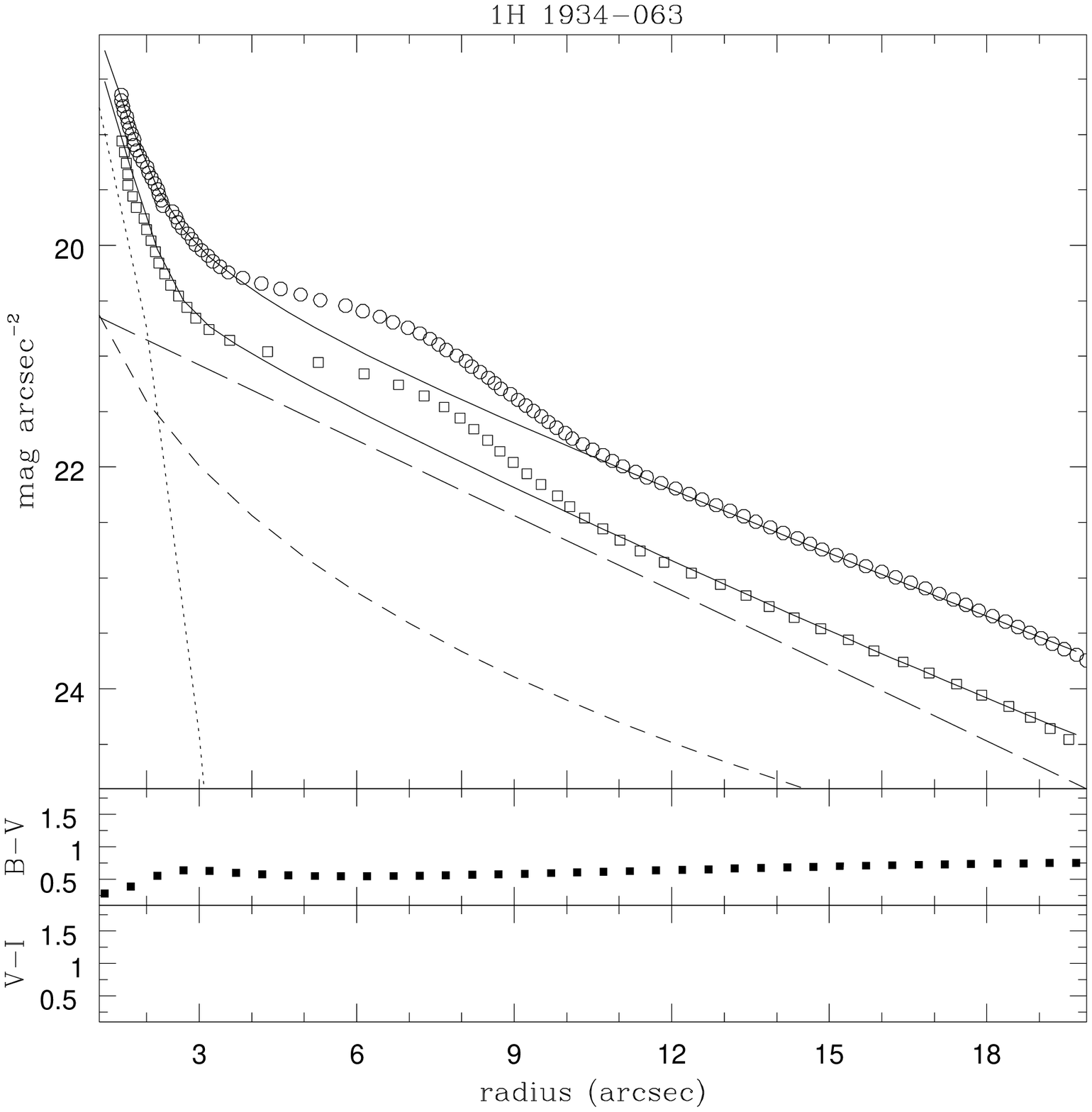}
	\end{figure*}

   \begin{figure*}
   \centering
   \includegraphics[width=8.7cm,height=8.7cm]{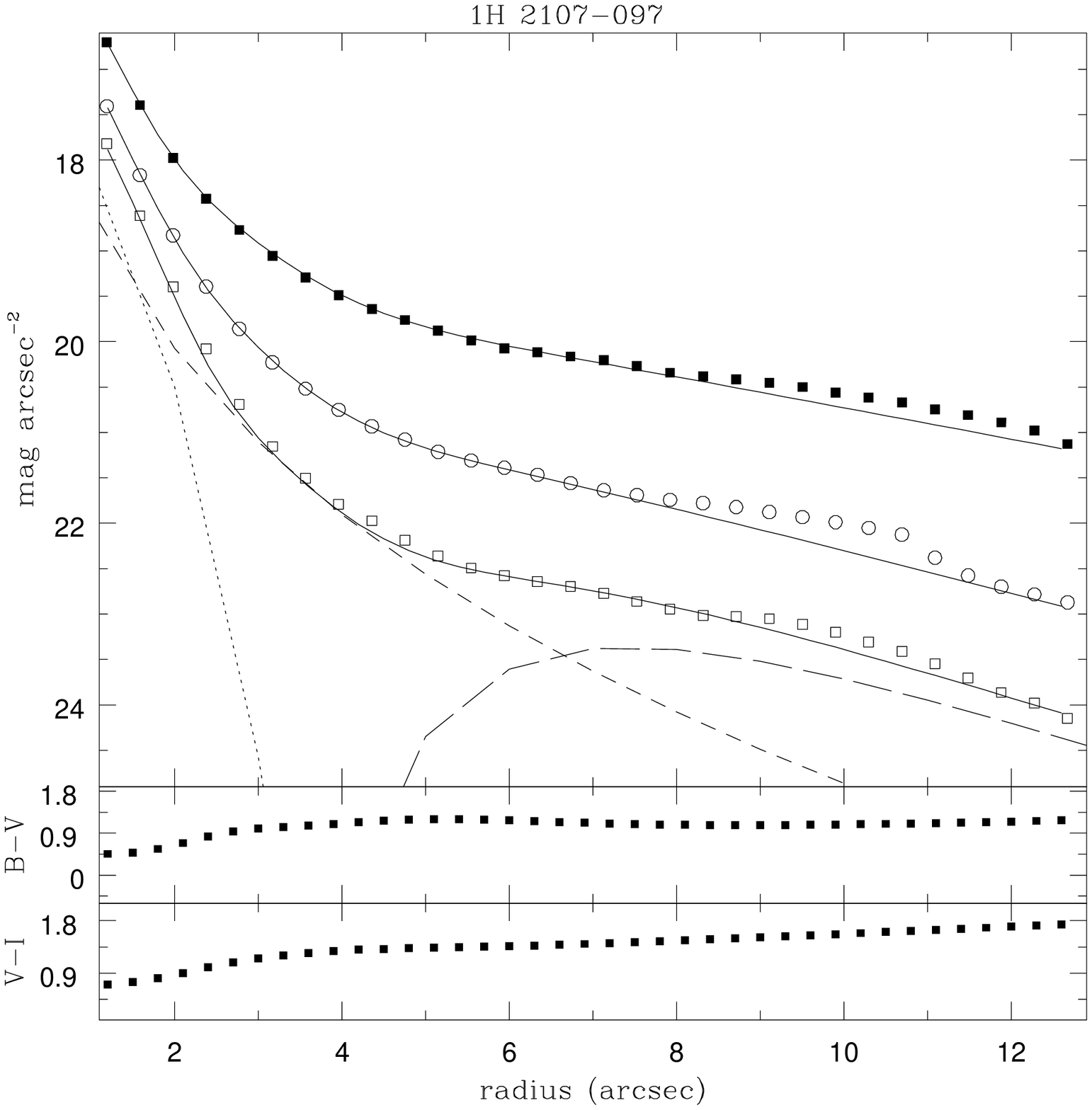}
   \includegraphics[width=8.7cm,height=8.7cm]{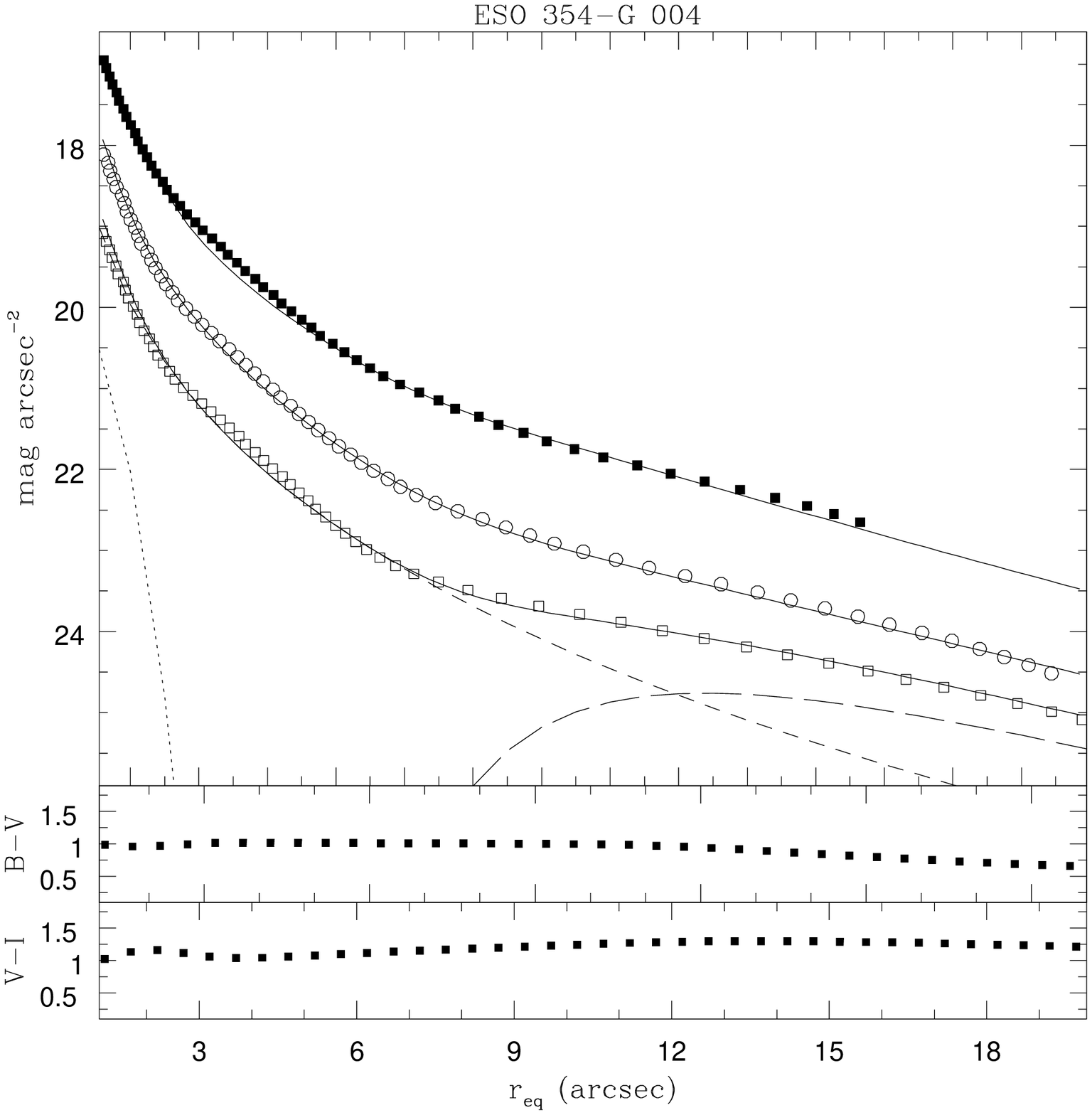}
   \includegraphics[width=8.7cm,height=8.7cm]{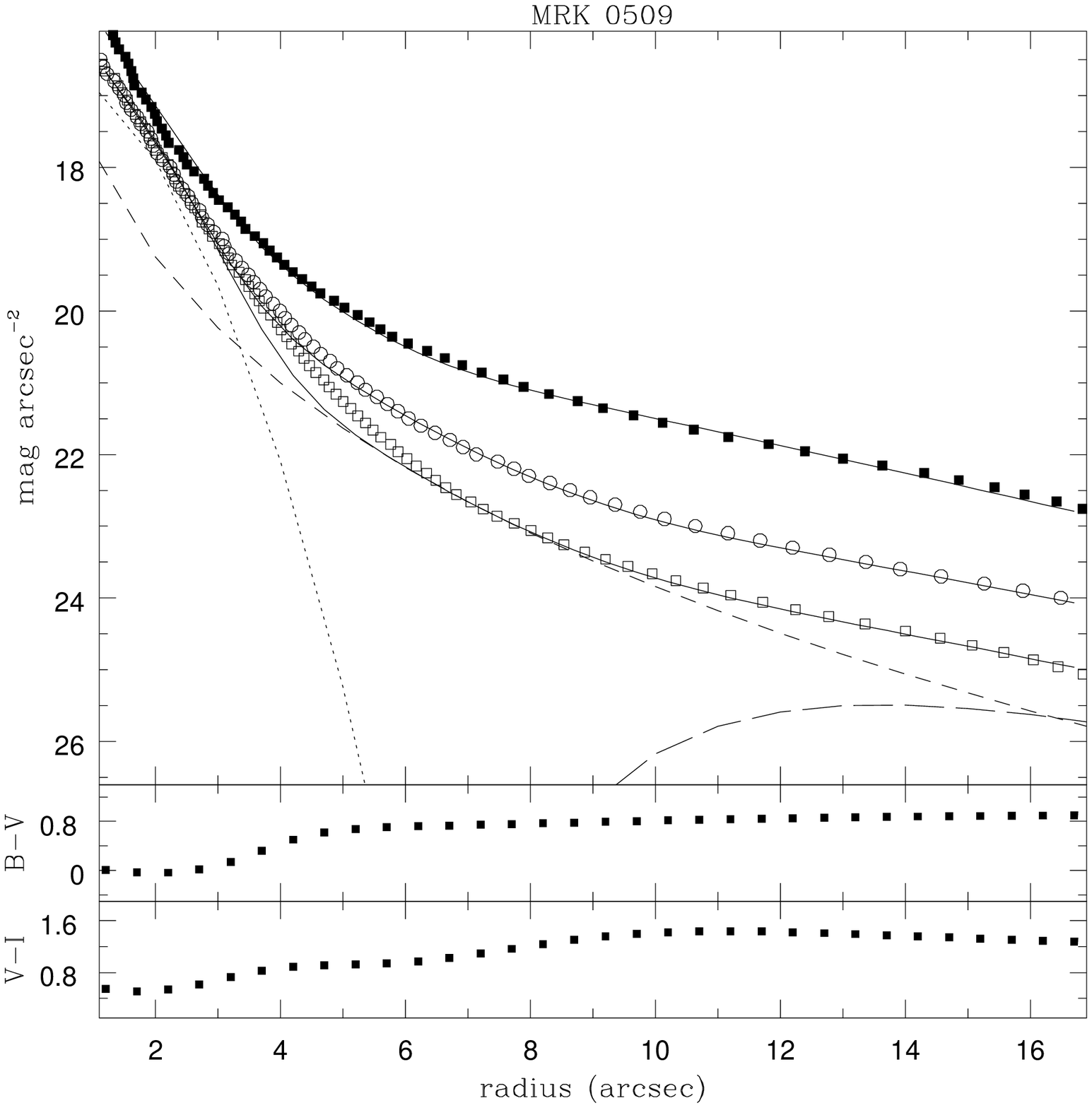}
   \includegraphics[width=8.7cm,height=8.7cm]{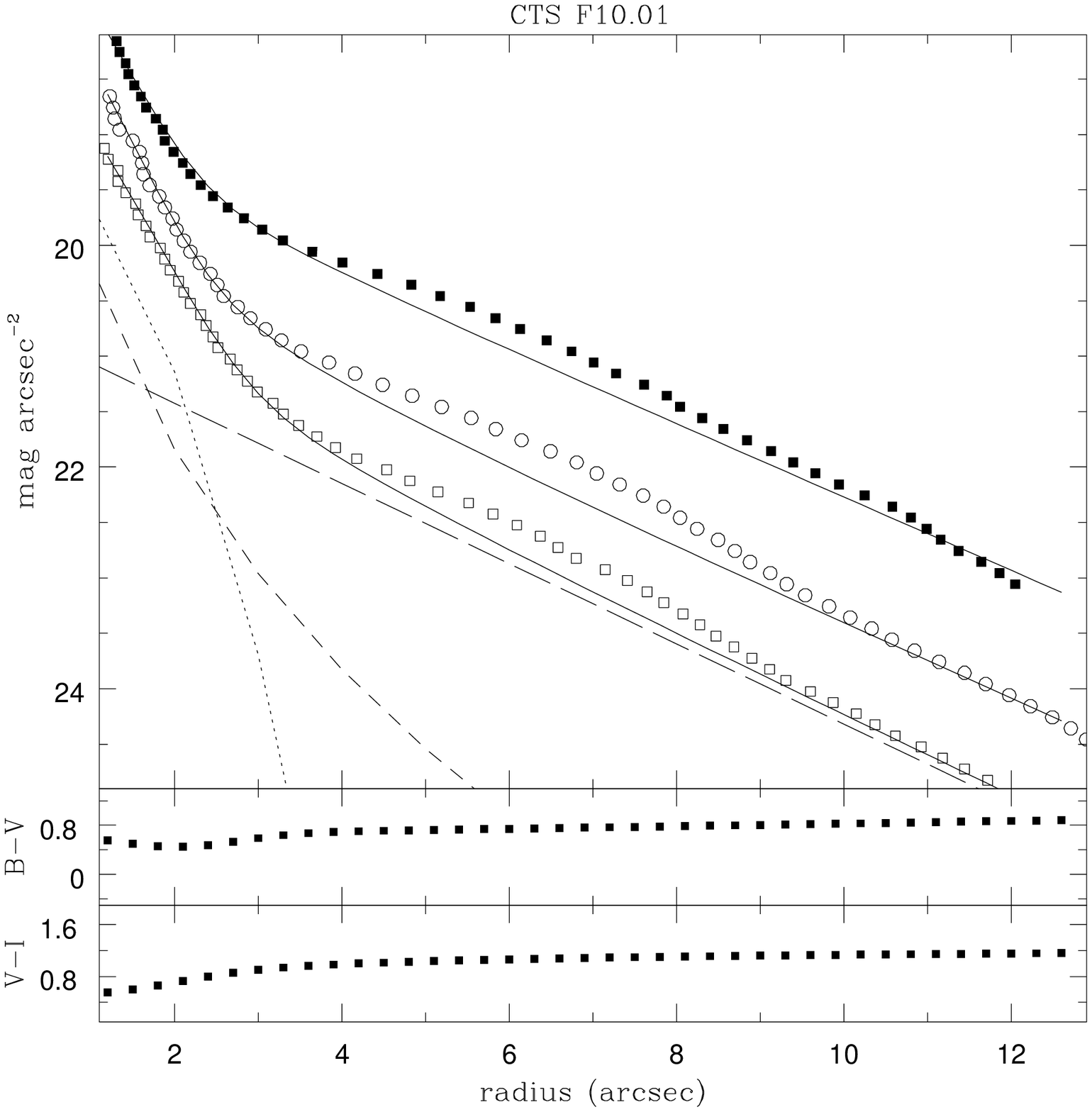}
  \caption{Luminosity B (empty squares), V (empty circles), 
I (filled squares) and color B-V and V-I profiles of the sample galaxies. 
Solid line is the best fit to the data. Decomposition in a gauss (points) 
bulge (short dashed) and disk components (long dashed) is only shown for
the B data. Color B-V and V-I profiles were calculated using the obtained 
fitted functions.
  a) CTC C16.16, b)CTS G03.04, c) CTS A08.12, d) ESO 602- G031,
  e) ESO 025- G 002, f)1H 1934-063, g) 1H 2107-097, h) ESO 354- G 004, i) MRK 
  0509, j) CTS F10.01.}
	\end{figure*}

\subsection{Color diagram and gradients}

The dominant stellar population of the individual galaxies is 
inferred from the integrated colors. This information, together with 
the luminosity profiles, can be used to derive the morphological type 
of the host galaxies. Figure 3 illustrates the color-color diagram 
(B-V) vs (V-I) for the galaxies listed in Table~\ref{morfo}. We have compared 
our data with those obtained from Hunt et al. (1999) for a sample of 
Seyfert 1 galaxies. The colors of both samples show a similar behavior. 
We have also plotted in that figure the integrated averaged colors of 
normal galaxies, taken from Table 2 of de Jong et al. (1996), for 
two different morphological types as well as Stellar population models 
for E galaxies with ages between 12 and 15 Gyr computed by Tantalo 
et al. (1998).

  \begin{figure*}
   \centering
   \includegraphics[width=8.7cm,height=8.7cm]{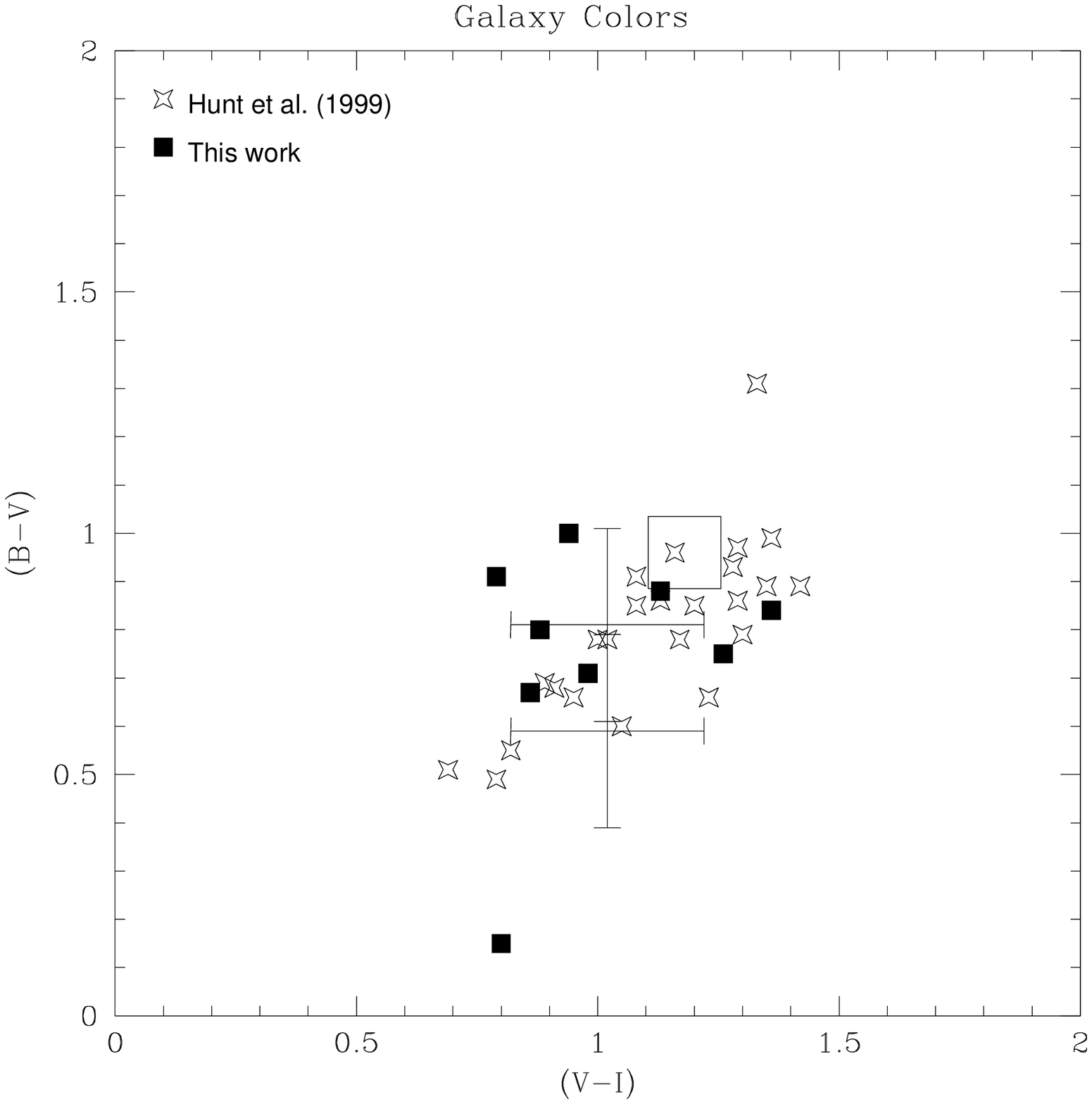}
\caption{Color-color diagram for the sample galaxies (filled squares).
Stars represent the Hunt et al. (1999) sample. Upper and lower crosses show the 
average color for normal galaxies with 0$<$T$<$2 and 6$<$T$<$8 respectively.
The small box show color models for elliptical galaxies with ages between 12 and 15
Gyr taken from Tantalo et al. (1998).}
	\end{figure*}

From the color$-$color diagram we conclude that: a) 40\% of the  
sample (ESO\,602-G\,031, 1H\,1934\,-\,063,
1H\,2107-097, and CTS\,F10.01) has a stellar population 
typical of Seyfert galaxies hosted by a spiral galaxy; 
b) CTS\,C16.16 and CTS\,A08.12 are early-type 
galaxies; c) The integrated (B-V) color of MRK\,509 is bluer than
that expected for any morphological type. However, it should be 
stressed that nearly half the luminosity of MRK\,509 comes from its 
AGN.

A quick inspection to the color profiles in Figures 2a-j
shows that almost all galaxies present color gradients.
In 1H\,1934-063, 1H\,2107-097, MRK\,509 and CTS\,F10.01, 
their B-V and V-I colors become monotonically redder with 
increasing distance to the galaxy center. In addition, the 
Seyfert nucleus leaves a clear signature in the color profiles: 
in the central 3$\arcsec$ colors are, on average, 0.3 mag bluer 
than those of the external parts of the galaxy. However,
ESO\,602-G\,031 and ESO\,354-G\,004 show redder B-V color gradients 
towards the center but their V-I color profiles are almost 
constant along the galaxy radius. It is worth 
noting that 1H \,2107-097 has a V-I color profile abnormally red 
(V-I $\sim$ 1.8) at large radius. 

Color profiles can be interpreted in terms of the spectral 
energy distribution of a given galaxy. Our sample is composed of
nine Seyfert 1 galaxies and one Narrow-Line Seyfert 1 galaxy. 
These type of galaxies have
strong emission lines and a continuum which can be accounted for by
a combination of stellar population and a non-thermal spectrum.
A significant difference in line intensity ratios and in the optical
spectral index is observed among Seyfert 1 and Narrow Line Seyfert 1 galaxies
(RPD2000). Seyfert 1s, like CTS C16.16, A08.12, CTS F10.01 
and  CTS G03.04  have continua steeper to the blue (RPD2000),
which is consistent with the observed  color profiles. The galaxy 1H 2107-097,
has a flat continumm (Figure 1c in RPD2000) and it is the only galaxy in our
sample that shows peculiar color profiles. The B-V profile is very blue in
the galaxy center while the V-I is abnormally red.

\subsection{H$\alpha$ images}

It is surprising to see that only one out of the 6 (16\%) imaged galaxies shows 
H$\alpha$ emission beyond the nucleus. In fact, we 
detected disk emission in ESO 025-G\,002 as it can be
seen from the continuum subtracted H$\alpha$ + [NII] image presented in Figure 4. 
From this image we can appreciate that emission is extended up to a distance of 
8 kpc from the nucleus.\\ 
The previous result is different from that reported by Pogge (1989)
(hereafter P89) who found
that 3 out of 9 (33\%) of the Seyfert 1 galaxies showed
extended emission within the inner 1 kpc. 
Moreover, P89 results contrast with those found by GD97
that report that 8 out of 13 (61\%) Seyfert 1 galaxies showed extended emission.
However, we must note that the mean distance for our sample is vz = 10500 km s$^{-1}$,
leading to a scale length of 0.7 kpc arcsec$^{-1}$. This is well below the resolution
of both P89 and GD97 works, for which the mean distance of the samples
is vz $\sim$ 2000 km s$^{-1}$ that leads to a scale length of 0.1 kpc arcsec$^{-1}$.\\
On the ohter hand, another important issue to have in mind is how 
these samples were defined. In principle,
the P89 and GD97 samples are similar, but while GD97 do not use interaction as
a selecting criterion, P89 excludes interacting galaxies. Similarly, P89 does not limit
the sample by inclination angle while GD97 do. None of these selecting critera
was used to define our sample. 

   \begin{figure*}
   \centering
   \includegraphics[width=8.7cm,height=8.7cm]{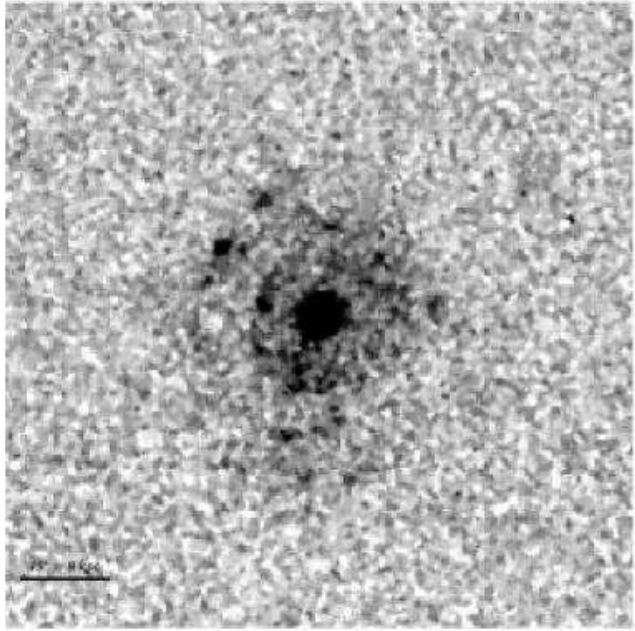}
  \caption{H$\alpha$ line emission image of ESO 025-G\,032. North is on top
   and East to the left. The lower left bar represents 15" and the corresponding projected
scale in Kpc is also indicated.} 
	\end{figure*}

  \section{The galaxies}
  In this section we describe the main photometric properties of
  the galaxies of the sample.

  \bigskip
  \noindent
  CTS C16.16:
  It appears in our images as
  elliptical (see Figure 1a). Its luminosity profile
  follows the $r^{1/4}$ up to 4$\arcsec$. Beyond this limit the
  profile shows an excess which may be suggestive of a merger
  scenario. In fact, we observe from our images a group of three small 
  galaxies with very disturbed isophotes towards the SW of CTS\,C16.16. 
  The biggest one, located to the W, is rather blue with (B-V) = 0.6.
  Color profiles show that outward of 3$\arcsec$ CTS\,C16.16 does not
  present a significant gradient, but inner to this radius both color 
  profiles become bluer. We interpret this result as due to the luminosity
  of the Gaussian component, which contributes to 30\% of
  the total luminosity. The integrated (B-V) and (V-I) colors
  are consistent with a dominant stellar population typical of an early 
  type galaxy. From the absolute blue magnitude of CTS\,C16.16 
  (M$_{\rm B}$=-20.36) and its dimension (r$_{\rm eff}$ = 2\,kpc) we conclude 
  that this object is a compact elliptical galaxy.

\bigskip
\noindent
CTS G03.04:
It is a ringed SO galaxy (see Figure 1b). 
Besides the ring, shells and plumes are also observed in our images. 
Despite its morphological classification, CTS\,G03.04 has rather 
blue integrated colors [(B-V)=0.80, (V-I)=0.88)], quite similar to 
those of an early spiral (Sa). We suggest that such colors may mostly 
be due to the AGN and disk luminosities. 
In fact, the blue disk luminosity contribution is almost a   
factor of 2 larger than that of the bulge and the AGN contributes with 
almost 20\% of the total galaxy luminosity. Color profiles show that 
the disk becomes important from 6$\arcsec$ outwards.
Inwards, the gradient becomes blue. The profiles also become very
blue towards the galaxy center, where the Seyfert nucleus dominates. 
A small galaxy, with (B-V)=0.81, is located at 31$\arcsec$ to the 
west of\,CTS G03.04.

  \bigskip
\noindent
CTS A08.12:
This object appears in our images as an E2 (see Figure 1c) but 
according to its luminosity profile, we classified it as SO due to
the presence of a disk that accounts for 50\% of the bulge luminosity.
In order to obtain a satisfactory fit to the luminosity profile 
the disk needed a cutoff radius at about 3 kpc.
The integrated colors are similar to a dominant stellar population of 
an E galaxy. The B-V profile shows a blue gradient that begins at 
r = 6$\arcsec$ outwards due to the presence of this disk. However, 
the V-I profile shows a red gradient with crescent radius. The most 
external isophotes of CTS A08.12 are rather disturbed probably by 
the presence of three small galaxies located at 9$\arcsec$, 
17 $\arcsec$ and 18$\arcsec$ to the NE from the nucleus.
The colors of these three objects are very similar, with (B-V)$\sim$1.7.
Towards the N, at about 50$\arcsec$, there is another SO
galaxy that shows very disturbed inner isophotes.
A closer inspection to the whole frame field (about 0.09 Mpc$^2$)
shows more than 20 galaxies brighter than m$_{\rm B}$=18.5
and with (B-V) colors ranging from 1.2 to 1.8. This evidence strongly
suggests that CTS\,A08.12 is located in a group or
a poor cluster, not yet identified in the literature.

  \bigskip
\noindent
ESO\,602-G\,031: 
This object  is a very  luminous  SABa (M$_B$= -21.28). It has
the typical integrated colors of normal spirals (see Figure 1d). 
The luminosity profiles are well fitted with the gauss + bulge + disk 
components. However, an excess above the fitted profile can be noted at 
r=5$\arcsec$ due to the presence of the bar, which is less prominent 
in the I band. The disk of ESO\,602-G\,031 shows a central cutoff 
radius of about 10 kpc. 
This feature is well noted in the B-V color map, which is presented in 
Figure 5a, evidencing 
that the inner 10 kpc are redder 
($\sim$0.3 mag) than the outermost regions.
The color profiles do not present very pronounced
gradients. The outer isophotes show evidence of
perturbation, suggesting that the galaxies located to the NE 
and SW are physical companions. Additional support to this
hypothesis can be drawn for the fact that these two galaxies
also show signs of tidal interactions.

  \bigskip
  \noindent
  ESO\,025-\,G002:
  According to its B image, we classify it as of
  SAB type. The galaxy is almost face-on, making evident not only the
  bar but also a ring located at 6 kpc from the nucleus (see
  Figure 1e). Along the ring, we detected several HII regions
  with fluxes lower than 10$^{-16}$ erg cm$^{-2}$ s$^{-1}$. It
  is interesting to note that this is the only galaxy of the sample
  that shows extended H$\alpha$ emission (see Figure 4). 
  The disk of ESO\,025-\,G002 needs a cutoff radius
  of $\sim$4kpc to properly fit the luminosity profile. The B-V color 
  map (see Figure 5b) and the B-V profile (see Figure 4b) clearly
  reveal this effect: a reddened inner region with r=5 kpc.
  The color of this region averages (B-V) = 1.2. 

  \bigskip
  \noindent
  1H\,1934-063:
  This galaxy is classified as elliptical in the RC3. However,
  it shows a very prominent disk and a rather blue color
  [(B-V)=0.61]. Moreover, several arms not well developed 
  are easily observed from our B image (see Figure 1f), 
  leading us to conclude that this objects is of Sb type. Its total radius
  (r=4 kpc) and its luminosity (M$_B$=-18.74) indicate that 
  1H\,1934-063 is a rather small galaxy. 
  The luminosity profiles shows an excess over the fitted profile
  that we interpreted as due to the presence of spiral arms.
  Luminosity profile decomposition shows that
  the disk is almost twice more luminous than the bulge. In the
  B band, the Seyfert nucleus contributes with 20\% of the total
  luminosity of the galaxy. No extended emission as well as
  possible companions are detected.

  \bigskip
  \noindent
  1H\,2107-097:
  We classified this galaxy as S0. From Figure 1g it
  can be seen a disk and a very faint halo that extents up to 18 kpc
  from the nucleus.
  However, it was not possible to trace the
  luminosity profile up to this radius due the bright star located
  to the W of the galaxy. The derived magnitude for this object up to a radius
  of 9 kpc is M$_{\rm B}$=-20.68, indicating a rather compact
  galaxy. The integrated color (B-V)=0.61 is typical
  of a normal spiral. In the B band, the Seyfert 
  nucleus contributes with 30\% of the total luminosity.
  The color profiles show steeper gradients towards the galaxy 
  center. However, in the outermost regions the B-V profile shows 
  a rather constant value while the V-I profile becomes abnormally 
  redder ($\sim$1.8). 1H\,2107-097 is another sample galaxy that 
  shows a disk with a cutoff radius. The
  value for this radius is $\sim$ 3 kpc.

  \bigskip
 \noindent
  ESO354-G\,004:
  This galaxy is cataloged in the RC3 as (R':)SA(rs)b. 
  However, no evidence of any ring structure is observed in Figure 1h.
  ESO\,354-G\,004 is
  rather luminous (M$_{\rm b}$=-21.22) and its radius extends up to
  20 kpc. The luminosity profiles are well fitted with a bulge and disk
  components. The disk shows a cutoff at about 10 kpc. This
  radius coincides with a red [(B-V)=1.1] region that can be observed from
its color map in Figure 5c. This region it is also noticeable 
  in the V-I profile. The integrated colors are consistent 
  with an early type stellar population. Signs of tidal interaction are 
  seen in the images. In fact, it is interesting to note that the 
  galaxy located towards the south seems to be connected with 
  ESO\,354-G\,004 by a very faint arm. On the other hand, the object 
  located towards the SW does not show any visible connection with the 
  main galaxy but the arms of ESO\,354-G\,004 are perturbed in 
  this direction. These two small galaxies have similar integrated
   colors, (B-V)$\sim$0.8.

   \begin{figure*}
   \centering
   \includegraphics[width=8.7cm,height=8.7cm]{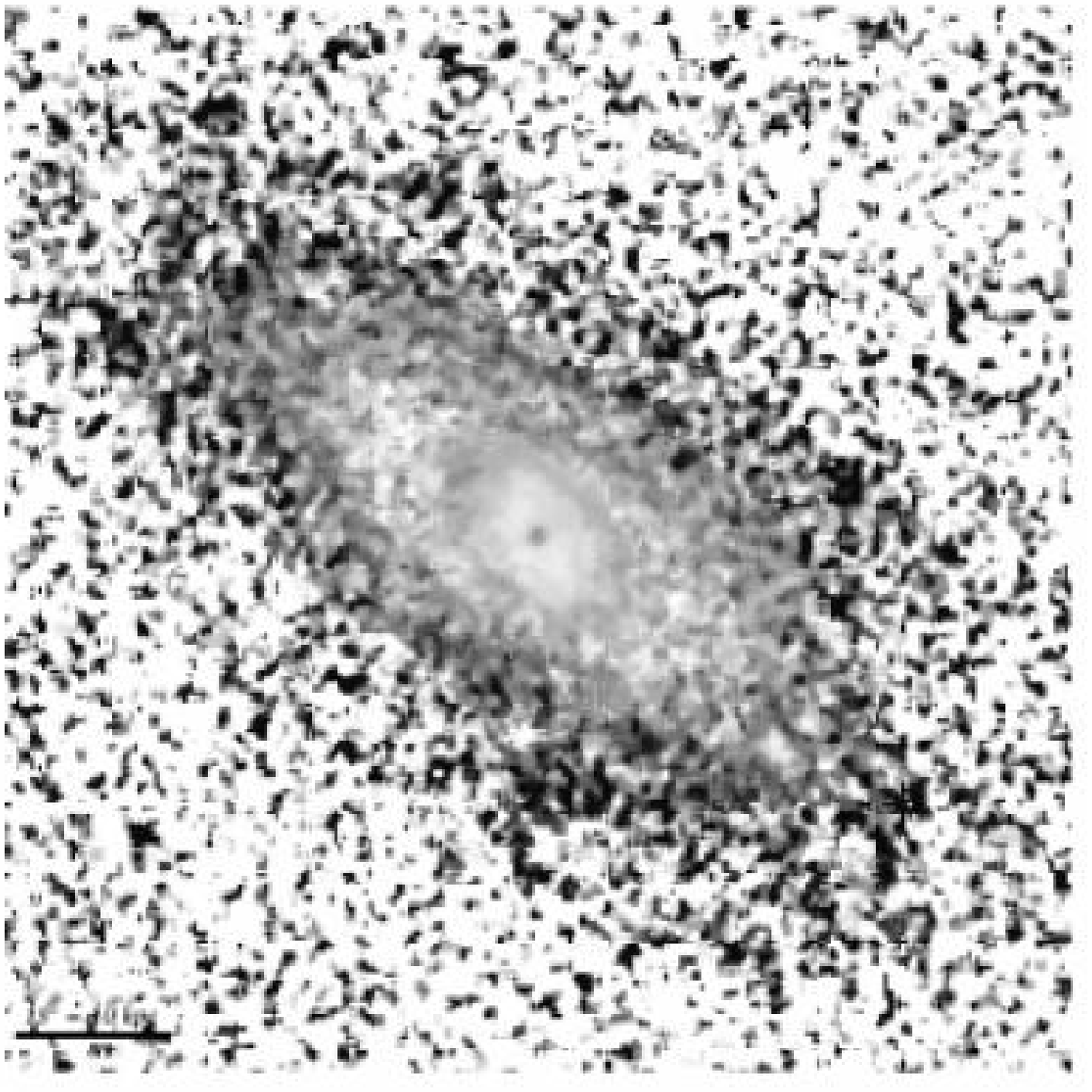}
   \includegraphics[width=8.7cm,height=8.7cm]{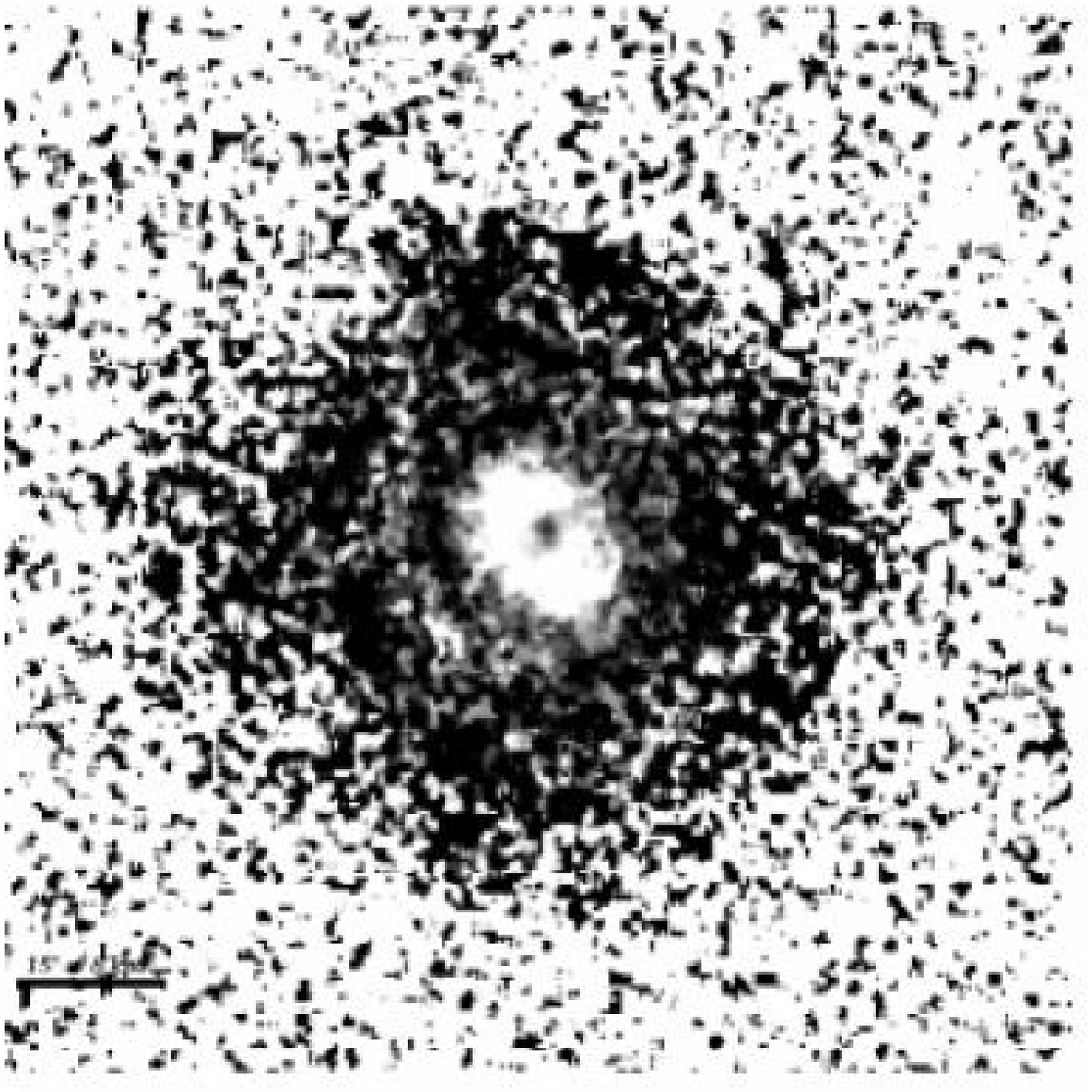}
   \includegraphics[width=8.7cm,height=8.7cm]{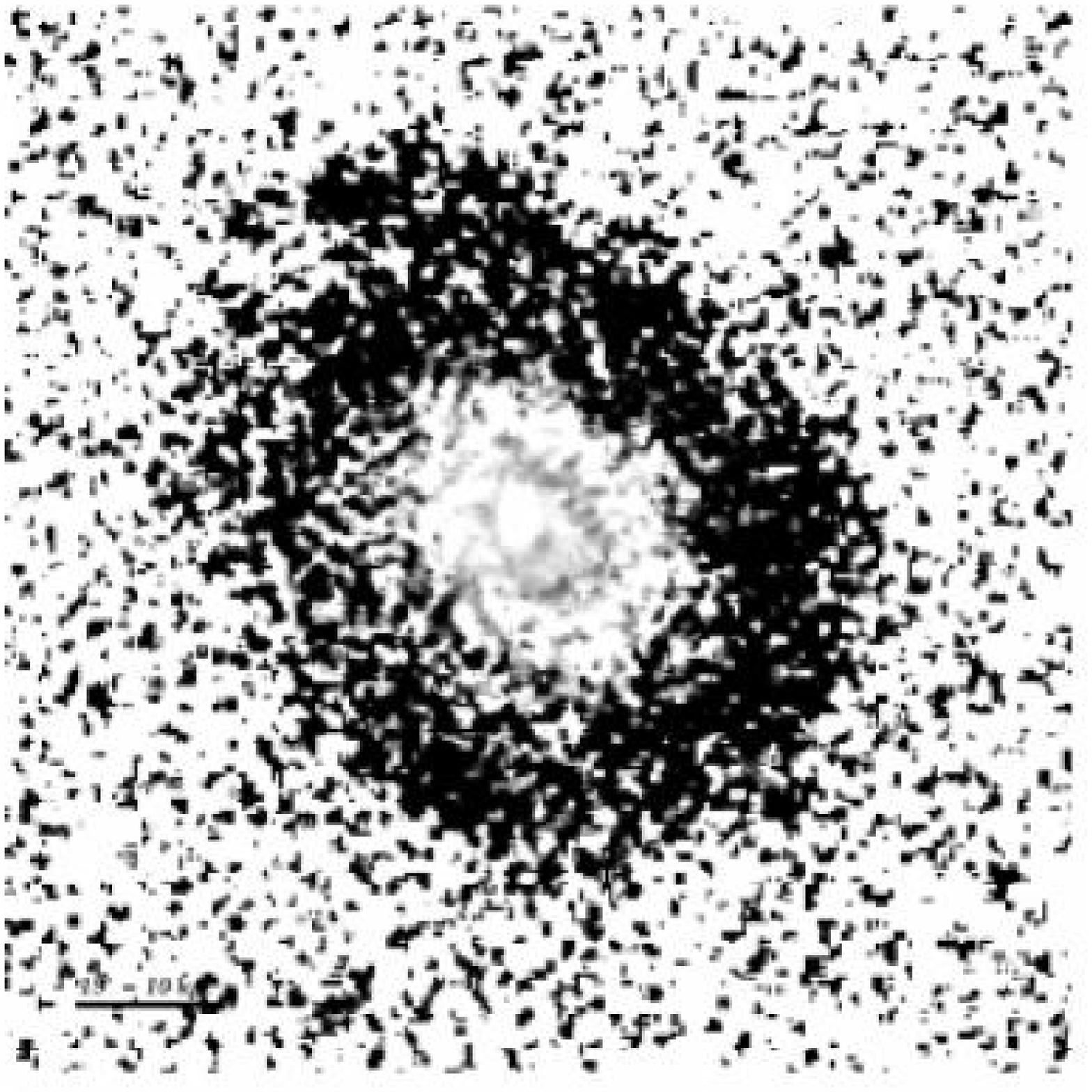}

  \caption{B-V color maps. North is on top and East to the left.
  Lower left bar represents 15" and the corresponding projected scale in Kpc
is also indicated. a) ESO 602- G031, b) ESO 025- G 002,
  c) ESO 354- G 004.}
	\end{figure*}
  
\bigskip
  \noindent
  MRK\,509:
This galaxy  is a very particular object (see Figure 1i). 
It is the most luminous and bluest galaxy [M$_{\rm B}$=-22.34, (B-V)=0.15] 
of our sample. In addition, it shows the largest luminosity contribution 
of the Seyfert nucleus to the total luminosity (L$_{gauss}$/L$_{total}$=0.56). 
This fact affects the integrated colors, which are abnormally blue compared
with colors of any morphological type. This object is quite well known
for being highly variable both in the continuum and emission lines
(Peterson et al. 1998). Its relatively small size
($\sim$10 kpc in radius) makes it also a rather compact object.
Apparently, it looks as an elliptical (E1). However, the luminosity
profile follows the r$^{1/4}$ law up to r=6.5 kpc. Beyond this limit, an
excess, which amounts 5\% of the bulge luminosity is well fitted with
a disk profile with a cutoff radius of 8.7 kpc. 
It is possible that the galaxy located towards the NW of MRK\,509 at about
3$\arcmin$ be a physical companion. In fact, it has a (B-V)=0.48 and shows
rather disturbed isophotes, giving additional support to this idea.

  \bigskip
  CTS\,F10.01:
  This galaxy is of S0 type and
  shows a prominent disk (see Figure 1j).
  From the luminosity profiles we derived a disk luminosity almost
  twice as large as that of the bulge. It has M$_{\rm B}$=-21.99 and a
  radius of 18 kpc, indicating that it is a compact object.
  Despite its morphological type, it is also rather blue [(B-V)=0.67].  
  Luminosity profiles show an excess respect to the fitting in
  the range 5$-$8$\arcsec$. This excess is similar in all bands and
  can be due to the presence of a lens structure. 

  \section{Conclusions}
  We have presented new photometric BVI data
  for 10 Seyfert 1 galaxies together with narrow band H$\alpha$
  images for 6 of these objects as well. The absolute B magnitudes
  of the galaxies, M$_{\rm B}$, are found to be
  spread over a large interval, from -18.74 to -22.34. Integrated 
  (B-V) and (V-I) colors as well as morphological types
  are derived for the first time in most of the objects.
  We found that the  morphologies  are confined to early type
  galaxies: one elliptical, five SO, one Sa and three Sb.
  Overall, 50\% of the objects can be considered as compacts. 
  Bars are found only in 2 cases
  (22\%). The (B-V) colors of the galaxies show to be biased to the
  blue. In fact, the SO galaxies of the sample shows, on
  average, (B-V)=0.78, significantly bluer than the
  average for this morphological type. We interpret this
  effect as due to a high contribution of the AGN and/or
  the disk to the total luminosity of the galaxy.

  Signs of tidal interactions are detected in six 
  galaxies of the sample. The case for CTS A08.12 is
  interesting since it seems to be located in a poor cluster not yet
  identified in the literature. However, it is not possible to confirm 
  if they are physical interacting objects due to the lack of radial
  velocities of the suggested companions. 

  Luminosity profiles were adequately fitted to the
  Gauss + bulge + disk components. In six out of eight
  disk models it was necessary to truncate the exponential profile 
  in order to improve the fit. The radius
  of the central cutoff ranged from 3 up to 10 kpc and it
  usually corresponds to reddened regions, generally well identified 
  in the B-V color maps. These regions present very similar colors 
  among the sample galaxies, (B-V)$\sim$1.2. We associated them 
  to the presence of dust in the inner few kiloparsecs of the 
  galaxies.

  The profile decomposition allowed us to derive the
  luminosity contribution of the AGN, bulge and disk separately. 
  We found that in the blue band the AGN contribution to the total
  luminosity varies from 3\% up to 56\%.
  In addition, the bulge to disk ratio ranges from 
  L$_{bulge}$/L$_{disk}$=0.6 to 22.

  H$\alpha$ images show that only 1 out of 6 galaxies presents
  disk emission. Additional data is needed in order to
  confirm if this emission is photoionized by the nuclear
  continuum or any starbutst component.

  \section{Acknowledgments}
We acknowledge Dr. S. Lumsden (referee) for his comments and sugestions.\\
This work was partially supported by SECyT,  PRONEX/FINEP grants 
76.97.10003.00 and Funda\c c\~ao de Amparo a Pesquisa do Estado de 
S\~ao Paulo $-$ FAPESP, under contract 00/01020-5.

\end{document}